\documentclass[traditabstract]{aa} 
\usepackage{graphicx}
\usepackage{txfonts}
\usepackage{textcomp}

\usepackage{natbib}
\bibpunct{(}{)}{;}{a}{}{,} 

\usepackage{txfonts}

\usepackage{rotating}
\usepackage{amssymb}
\begin{document}
  \title{Mining SDSS in search of Multiple Populations in Globular Clusters.}


   \author{C. Lardo
          \inst{1}, M. Bellazzini\inst{2}, E. Pancino\inst{2}, E. Carretta\inst{2}, 
	  A. Bragaglia\inst{2}, \and E. Dalessandro\inst{1} 
          }
  
  \authorrunning{Lardo et al.}

   \institute{Department of Astronomy, University of Bologna, Via Ranzani 1,
 40127 Bologna, Italy\\
             \email{carmela.lardo2@unibo.it}
         \and
             INAF-Osservatorio Astronomico di Bologna, Via Ranzani 1,
 40127 Bologna, Italy\\
     }

   \date{Received/Accepted}

 
  \abstract
    {Several recent studies have reported the detection of an anomalous color 
spread along the red giant branch (RGB) of some globular clusters (GC) that appears
only when color indices including a near ultraviolet band (such as 
Johnson {\it U} or Str\"omgren {\it u}) are considered. This anomalous spread in
color indexes such as {\it U-B} or {\it $c_y$} has been shown to correlate with
variations in the abundances of light elements such as C, N, O, Na, etc., which, in
turn, are generally believed to be associated with subsequent star formation
episodes that occurred in the earliest few $10^8$~yr of the cluster's life.
Here we use publicly available {\it u}, {\it g}, {\it r} Sloan Digital Sky 
Survey photometry to
search for anomalous {\em u-g} spreads in the RGBs of nine Galactic GCs. In seven of them 
(M~2, M~3, M~5, M~13, M~15, M~92 and M~53), we find evidence of a statistically
significant spread in  the {\it u-g} color, not seen in {\it g-r} and not
accounted for by observational effects. In the case of M~5, we demonstrate that
the observed {\it u-g} color spread correlates with the observed abundances of Na, 
the redder stars being richer in Na than the bluer ones. In all the seven clusters
displaying a significant {\it u-g} color spread, we find that the stars on the
red and blue sides of the RGB, in ({\it g, u-g}) color magnitude 
diagrams, have significantly different radial distributions. In particular, the
red stars (generally identified with the second generation of cluster stars, in
the current scenario) are always more centrally concentrated than blue stars
(generally identified with the first generation) over the range sampled by the 
data ($0.5r_h\la r\la 5r_h$), in qualitative agreement with the
predictions of some recent models
of the formation and chemical evolution of GCs.
Our results suggest that the difference in the radial distribution between first
and second generation stars may be a general characteristic of GCs.}

\maketitle

\section{Introduction}
\label{intro}
Globular clusters (GCs) have long been considered as examples of the theoretical concept of a simple 
stellar population, i.e., a population of stars that is strictly coeval and chemically homogeneous \citep{renzini}.
This traditional paradigm remains valid in some cases, but recent photometric
and spectroscopic observational results strongly indicate that most GCs have been sites of two or even 
more star-formation episodes, producing a peculiar chemical (self-)enrichment pattern
\cite[see][and references therein, for review and discussion]{piotto2009,car10}.\\

In general, GCs are largely homogeneous if we consider Fe-peak elements but exhibit a significant spread
in the abundance of lighter elements, with strong anti-correlations between, for example, the 
abundances of Na and O, or Mg and Al, as well as bimodal distributions of CH and CN line strengths 
\cite[see][and references therein]{car10,mart09}.
While these anomalies have been known for decades \citep[see][for a review of early results]{kraft94}, 
it has only recently been established that (a) they can also be traced in un-evolved stars 
\citep{gratton01,ramirez02,carretta04}, hence cannot be due to mixing processes, and (b) they seem to be a characteristic
feature of GCs \citep{car10}.
Interestingly enough, these chemical inhomogeneities are not confined only to Galactic GCs \citep[][and references therein]{colucci09,mucciarelli09,johnson06,letarte06}.\\

The scenario generally invoked to explain the above phenomena,
foresees two subsequent generations of stars in the first few hundred Myr of cluster life, 
with ejecta from the massive stars of the first generation enriching the intra cluster medium (ICM) before the 
formation of the second one. Intermediate-mass asymptotic giant branch stars \citep[IM-AGBs;][]{ventura2008} or 
fast rotating massive stars \citep[FRMSs ;][]{decressin} have been proposed as the ICM 
polluters that most likely explain the observed patterns \citep[see][for a discussion of the merits of the two models]{alvio08}.

In this context, several results imply that the 
spread in light-element abundances of GCs can be traced by photometric indices with near 
ultraviolet (wide or intermediate) passbands, encompassing the wavelength range 3000\AA $\la\lambda\la$ 4000\AA. 
For instance, \citet{yong08} demonstrated that the  $c_{y}$ index 
\citep[including the near-ultraviolet Str\"omgren {\it u} passband, akin to the {\it $c_{1}$} index used by]
[to trace NH\footnote{In particular, $c_y=c_1-(b-y)$ and $c_1=(u-v)-(v-b)$, see \citet{yong08}, 
and references therein.}]{gru01,gru02} very clearly traces the differences in N abundances among the stars 
of NGC~6752, over a range of $\sim 2.0$ dex in [N/Fe]. As the abundance of N is found to correlate 
(or anti-correlate) very well with other light elements exhibiting abundance spreads in GCs (O, Na, Mg, Al), 
differences in these elements can also be correlated with {\it $c_{y}$} variations, as demonstrated 
by \citet{car09b}. These authors used their Na and O abundance determinations for 
NGC~6752 to show that stars with different abundances of these elements 
lie on  different sides of the cluster RGB in the ({\it V,$c_{y}$}) plane, and concluded that 
Na-poor (Na-rich) stars are also N-poor \citep[N-rich; see also][for further results on NGC~6752 
and discussion]{milone2010}. Moreover, the well-known anti-correlation between Na and O
implies that Na-rich stars are also O-poor and vice versa.
\citet{yong08} noted that available Str\"omgren photometry of Galactic GCs \citep{gru99} 
suggests that a significant spread in N abundance may be a general property of GCs. 
In particular, their Fig.~14 shows remarkable spreads in $c_{y}$, at any magnitude along the RGB of NGC~288, 
NGC~362, M~5, M~3, M~13, NGC~6752, M~15, M~92, and NGC~6397.

\citet{marino2008} were able to demonstrate for M~4 that a bimodal Na-O distribution, correlated  
with a bimodal distribution in CN strength, was also clearly associated with a bimodal spread in 
the color of RGB stars  in the {\it U} vs. {\it (U-B)} CMD, not seen with other color indices.
In this case, as in all cases in which the {\it U--B} color spread in the RGB has been 
correlated with O, Na abundances, the Na-rich/CN-strong stars lie on the red side 
of the RGB, while Na-poor/CN-weak stars populate a bluer portion of the branch.
In line with \citet{yong08}, \citet{marino2008} attribute the spread in 
{\it (U-B)} color to strength variations of 
several NH and CN bands included within the {\it U} passband, in particular the NH band around 
3360\AA, and the CN bands around 3590\AA, 3883\AA ~and 4215\AA. They performed synthetic 
{\it U}, {\it B} photometry of theoretical spectra with different strengths of the NH and CN features, 
mimicking the observed spectra of M~4 stars, finding that the resulting variation in the 
{\it U-B} color goes in the right direction but has an amplitude four times smaller than the observed color spread.

\citet{kravtsov10b,kravtsov10a} used {\it U} band photometry to identify 
multiple populations within the clusters NGC~3201 and NGC~1261. The spread in {\it U-B} color among RGB stars of NGC~3201 was 
shown to correlate with Na abundance by \citet{car10}.
Finally, \citet{han2009} showed that the RGB of NGC~1851, which is narrow and well defined in optical 
CMDs not including {\it U} photometry, is clearly split into two parallel branches in the {\it U} vs. {\it U-I} colour. 
\citet{han2009} suggest that the splitting is caused by a combination of effects 
due to variations in the abundances of not only of C,N,O but also of heavier elements 
(Ca, Si, Ti, and Fe) and helium. The presence of a small (but real) spread in iron and $\alpha$-elements abundance
 in this cluster was recently confirmed spectroscopically by \citet[][see also their Fig.~4, where they show that 
the cluster stars segregate along the RGB according to their Na abundance as in the other cases described above]{eu1851}.
Additional results and a more general discussion about 
the possible role of heavy elements variations in producing the RGB color spreads in GCs can be found in \citet{leenat} and \citet{carretta2010}.\\

While the details of the process remain still unclear, it seems reasonably well established that
(at least) light-element abundance spreads in GCs are correlated with spreads in the colors of 
RGB stars if near ultraviolet filters are adopted. This opens an interesting window 
in investigating the origin of the chemical evolution of GCs as accurate {\it U} photometry for very 
large sample of GC stars can be obtained in a much easier way and for much more distant clusters  
than the mid-to-high-resolution spectroscopy needed to obtain direct 
chemical abundance estimates (which, however, provides more information). In particular, wide-field photometry 
capable of discriminating between N-rich and 
N-poor stars would provide the large samples and the wide radial coverage that are needed 
to compare the radial distributions of the two groups; this may provide a very interesting insight 
into the process of GC formation and early chemical enrichment 
\citep[see][]{dercole08,decressin,alvio08}.

Prompted by these considerations we decided to test whether additional observational evidence of 
color spreads in the RGB of GCs could be derived from the publicly available Sloan Digital 
Sky Survey photometry \cite[SDSS; see][and references therein]{dr7}, as it provides multicolor 
photometry including a {\it u} passband\footnote{The SDSS {\it u} filter has $\lambda_{eff}$=3521\AA~and 
FWHM=555\AA. For details about the photometric filters quoted in this paper see the Asiago 
Database of Photometric Systems {\tt http://ulisse.pd.astro.it/Astro/ADPS/} \citep{adps}.} 
and incorporates several Galactic GCs. In the following, we use the accurate {\it ugriz} SDSS 
photometry by \citet{an2008} of nine GCs selected from their sample to show that: (a) the {\it u-g} color 
index indeed correlates with spectroscopic Na,O abundances and that the associated 
spread in color can be detected in SDSS data, at least in clusters of intermediate to 
high metallicity; (b) a statistically significant intrinsic {\it u-g} spread among RGB stars is detected 
in seven of the nine clusters considered in the present study; and, finally, (c) in these 
seven clusters the radial distribution of stars
lying on the blue side of the RGB (in {\it u-g}) differs significantly from that of stars lying on the red side, the latter being more centrally concentrated than the former. 

In Sect.~\ref{data}, we briefly describe the adopted photometric data set and present 
evidence of a correlation between light-element abundances and {\it u-g} spread along the RGB for the cluster M~5.
In Sect.~\ref{photo}, we analyze the {\it g,g-r} and {\it g,u-g} CMDs of the considered clusters in search of
{\it u-g} color spread in the RGB that cannot be accounted for by observational effects. 
We finally discuss our results in Sect.~\ref{conclusion}.
Some preliminary results from this study were presented by M. Bellazzini at the meeting 
''The Giant Branches'' held in Leiden in May 2009
\footnote{\tiny\tt www.lorentzcenter.nl/lc/web/2009/324/Friday/Bellazini.ppt}.\\

\begin{figure}
  \centering
  \includegraphics[width=8cm]{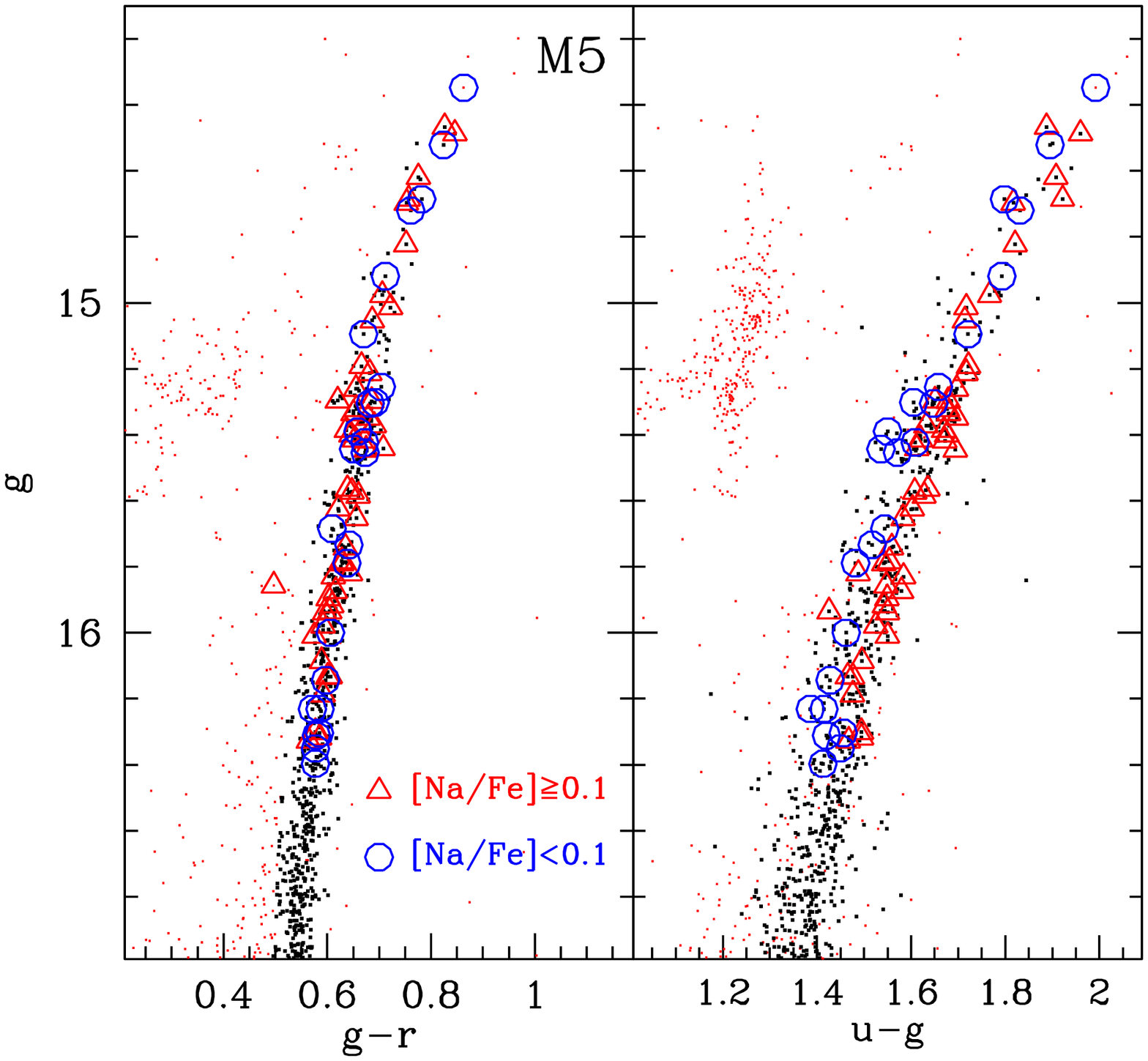}
  \includegraphics[width=8cm]{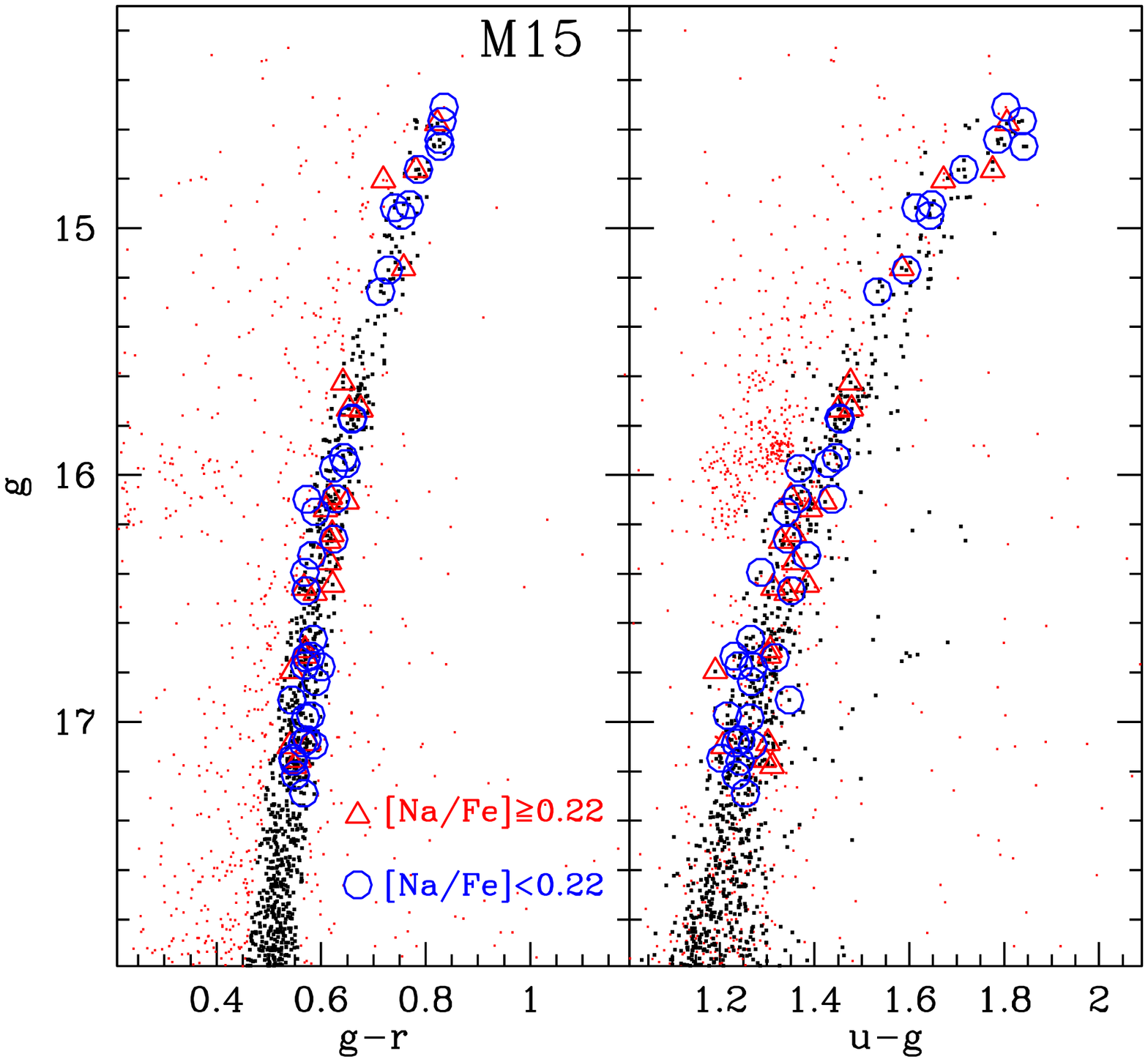}
\caption{ {\it g,g-r} and {\it g,u-g} CMDs for M~5, M~15. The stars for which spectroscopic abundances 
are available from from \citet{carretta2009,car09b} are marked according to their Na abundances: 
Na-poor stars 
are marked as blue open circles, while Na-rich stars 
are marked with red open triangles. Stars plotted as heavy dark dots are those selected as candidate 
RGB stars from the {\it g,g-r} CMD. 
             }
        \label{Spectro}
   \end{figure}

\section{The data set}
\label{data}

\citet[][An08 hereafter]{an2008} reanalyzed SDSS images of the GCs (and open clusters) included in the survey using 
the DAOPHOT/ALLFRAME suite of programs \citep{stetson1987, stetson1994}. These programs are known to perform more effectively
than the survey photometric pipeline in the high-crowding 
conditions typical of dense star clusters. 

We limited our analysis to the most favorable cases, excluding clusters with $|b|\le 20\degr$ - to avoid strong contamination 
from Galactic field stars - and/or with $E(B-V)\ge 0.15$ - i.e., clusters with relatively high extinction. Moreover, we 
decided to include in our sample only clusters with more than 100 candidate RGB stars 
between the horizontal branch (HB) and two magnitudes below this level, to ensure a solid statistic basis to 
our analysis. According to these criteria, 
among the seventeen clusters considered by An08 we selected the nine GCs listed in Tab.~\ref{GCs}. 

Only stars with valid 
magnitude estimates in both g and r have been retained for subsequent analysis. When An08 provided more
that one catalog per cluster 
\citep[i.e., in cases of clusters  imaged in different overlapping SDSS ``plates'', see][An08, and references therein]{dr7},
we merged the catalogs into a single one including all the cluster stars 
listed by An08 without duplications. The final catalogs cover fields including the vast majority 
of cluster stars but does not reach the tidal radius (except for NGC~2419, see below).

In some cases, the available photometry was highly 
incomplete and/or was characterized by large uncertainties in the innermost regions of the 
cluster, because of the high degree of crowding. In most cases the final catalog also covered 
a field of view including regions where the contamination from Galactic stars was not negligible. To avoid the 
inclusion of cluster stars with poor-quality photometry and contamination from fore/background 
Galactic stars, we limited the analysis to a radial corona between the $r_{in}$ and $r_{out}$ values specified in Tab.~\ref{GCs}.
In this table, $r_{in}$ and $r_{out}$ are also expressed in units of core, half-light, and tidal radii 
($r_c$, $r_h$, and $r_t$, respectively), to provide a more objective idea of the actual 
radial sampling and of the cluster-to-cluster differences in the radial sampling. We note that: (a)
in all the considered cases we sample the clusters out to  $>4~r_h$, and in most cases
out to $6-10~r_h$; (b) in the cases in which we excluded the innermost $1\arcmin$ from the analysis, the innermost
regions between $\sim 1-3~r_c$ were lost, the completeness in these inner parts possibly being
remarkably low also in clusters for which we retained the central region\footnote{This should not affect 
the results presented below, as in all the cases we compare RGB stars having the 
same distribution in magnitude, see Sect.~\ref{radsec}.};
and (c) the considered clusters span a wide range of central density \citep[more than 4 
orders of magnitude, see][]{harris1996}, hence the different radial ranges also sample regions in widely 
different dynamical conditions, depending on the considered cluster.

The reddening for the GCs are taken from \citet{harris1996}. 
The coefficients of the adopted extinction laws ($A_{\lambda}/A_{V}$) 
are taken from the computations by \citet{girardi04} for cool giants ($T_{eff}$=4000, $\log$~\textit{g}=2.00 
and [M/H]=-2), and assuming $A_V=3.1E(B-V)$. In particular:
$A_u=4.84E(B-V)$, $A_g=3.64E(B-V)$, and $A_r=2.71E(B-V)$, and $E(u-g)=1.20E(B-V)$ and 
$E(g-r)=0.93E(B-V)$. 
\begin{table*}
\caption{Globular cluster sample}             
\label{GCs}
     
\centering          
\begin{tabular}{l c c c c c c c c c}
\hline\hline

NGC& Alt. Name & $r_{in}$ &$r_{out}$ &$r_{in}/r_c$&$r_{in}/r_h$ &$r_{out}/r_h$ &
$r_{out}/r_t$&E(B--V)  & [Fe/H]  \\
   &           & arcmin   & arcmin   & & & & &         &         \\   
\hline                    
   2419&  	    & 0.0 	& 7.0  & 0.0&  0.0&  7.3&  1.0& 0.11 &  --2.12	  \\  
   5024&  M~53  	& 1.0 	& 7.0  & 2.8&  0.9&  6.3&  0.3& 0.02 &  --1.99	  \\
   5272&  M~3  	& 0.0  	& 10.0 & 0.0&  0.0&  8.9&  0.3& 0.01 &  --1.57	  \\
   5466&  	    & 0.0  	& 10.0 & 0.0&  0.0&  4.4&  0.3& 0.00 &  --2.22	  \\
   5904&  M~5  	& 0.0  	& 10.0 & 0.0&  0.0&  4.7&  0.3& 0.03 &  --1.34	  \\
   6205&  M~13  	& 1.0  	& 10.0 & 1.3&  0.7&  6.7&  0.4& 0.02 &  --1.54	  \\
   6341&  M~92   & 0.0   & 10.0 & 0.0&  0.0&  9.2&  0.7& 0.02 &  --2.28 \\
   7078&  M~15  	& 0.0 	& 10.0 & 0.0&  0.0&  9.4&  0.5& 0.10 &  --2.32	  \\
   7089&  M~2  	& 1.0  	& 10.0 & 2.9&  1.1& 10.7&  0.5& 0.06 &  --1.62	  \\
\hline 
\end{tabular}
\tablefoot{Global parameters are from the most recent version of the 
\citet{harris1996} on-line database (year 2003), 
except for [Fe/H] for both M~5 and M~15, which are taken from \citet{carretta2009}, and structural parameters for NGC~2419 that are taken from \citet{mic24}.}
\end{table*}

\subsection{Na abundances and {\it u-g} colors in RGBs}
\label{data-spec}
For two of the selected clusters, M~5 and M~15, we found a significant sample of stars for which 
there are spectroscopic Na abundances over a remarkably wide luminosity range along the RGB, from 
\citet{carretta2009, car09b}.
Following the criteria used by these authors, we divide the RGB stars between a
candidate first generation and a candidate second generation 
(P and I+E components, respectively, adopting their 
nomenclature\footnote{P stands for the first - {\em Primordial} - generation (Na-poor stars), 
while Na-rich stars are divided into I ({\em Intermediate}) and E ({\em Extreme}) 
subsequent generations. For homogeneity, we follow the nomenclature by Carretta et al. but  
we always consider all
Na-rich stars as a single class (I+E), for simplicity.}) by adopting a threshold
in sodium abundance that depends on the cluster metallicity. In particular, stars having
$[Na/Fe]_{min}\le [Na/Fe]< [Na/Fe]_{min}+0.3$ are assigned to the Na-poor P component, while stars having $[Na/Fe]\ge [Na/Fe]_{min}+0.3$ are assigned to the Na-rich 
I+E component \citep[see][for further details]{car09b}. The resulting thresholds are
[Na/Fe]= +0.10 for M~5 and [Na/Fe]= +0.22 for M~15; we stress, however, that the results presented below are not particularly sensitive to the adopted threshold.

The upper panel of Fig.~\ref{Spectro} shows that, in the case of M~5, 
{\em the Na-poor and Na-rich stars, that are tightly aligned along the narrow cluster RGB 
in the {\it g,g-r} CMD, are clearly separated into two parallel sequences in the much broader giant 
branch seen in the {\it g,u-g} diagram, the Na-rich stars appearing systematically redder 
than Na-poor ones} (at least below the HB level), a behavior strictly analogous to that observed by \citet{marino2008} 
in M~4, by \citet{milone2010} in NGC~6752, using {\it U} photometry,
and by \citet{car09b}, again in NGC~6752, but using Str\"omgren $u$ photometry.

This result clearly 
indicates that SDSS {\it u} photometry can also be used to trace the UV color spread correlated with 
the light-element abundance spread described in Sect.~\ref{intro}, above. We note that 
most of the RGB stars displaying the 
color segregation as a function of Na abundance {\em are fainter than the RGB bump} (at $g\simeq 15.4$, see Fig.~2, below), 
as expected if the observed chemical anomalies are not due to extra-mixing phenomena
known to occur in evolutionary phases brighter than this 
feature \citep{grat2000,smith03}.

However, the clear {\it u-g} color segregation between Na-poor and Na-rich stars observed in 
M~5 is not seen in M~15 (lower panel of Fig.~\ref{Spectro}). A likely explanation of 
this different behavior calls into play the difference in overall metal content between 
the two clusters. The iron abundance in M~5 is ten times higher than in M~15 and the 
average abundance of the light elements should scale similarly. Hence, the same degree 
of N abundance with respect to iron (as expressed by [N/Fe]) corresponds to very 
different absolute abundances of N ([N/H]). This, in turn, should correspond to 
significant differences in the strength of the absorption features that are supposed 
to drive the spreads observed in broad-band near-ultraviolet photometry 
\citep[see also][]{martell08}. In particular, the strengths of spectral lines of diatomic molecules (such as CN, NH) 
depend quadratically on the overall metallicity and it has been noted that CN bands become very weak in 
the spectra of GC giants for [Fe/H]$ \la -1.8$ \citep{smith02}. In this context, 
we note that the clusters for which the color spread in the RGB has 
been detected using the broad {\it U} (or {\it u}, in the present case) filter and correlated with spectroscopic 
light-element abundances, have 
intermediate metallicities, i.e. 
[Fe/H]=--1.34 for M~5, [Fe/H]=--1.17 for M~4, [Fe/H]=--1.51 for NGC~3201, [Fe/H]=--1.55 for NGC~6752
\citep[metallicities from][]{car09b}, and 
[Fe/H]=--1.18 for NGC~1851, \citep[metallicity from][]{eu1851}.\\

On the other hand, we need also consider that the RGBs of M~5 and M~15 seem to 
display similar spreads in the ({\it V,$c_{y}$}) plane \citep{yong08}, thus not 
supporting the above hypothesis. It is difficult to draw a firm conclusion at the present stage, 
in particular if we consider that we still lack a detailed theoretical understanding of 
the mechanism at the origin of the color spread. This kind of investigation is clearly 
beyond the scope of the present paper. For our present purposes, we take as the basis of 
the further analysis of the clusters in our sample the clear evidence that, at least in some cases, star-to-star 
differences in light-element abundances in the RGB of GCs can be discerned with the $u-g$ color.

In particular, the sought-after anomaly should be distinguishable as a significant color spread in the RGB in 
the ({\it g,u-g}) CMD, not seen in the ({\it g,g-r}) CMD
and unaccounted for by other effects, such as an increase in the photometric error and/or an increase in the amplitude of 
the effect of differential extinction (see below). 
Independently of its true origin, we refer in the following section to the observational 
effect under consideration as a {\em UV-spread} (UV-s hereafter, for brevity) and we will 
look for its presence in the clusters of our sample, trying to establish, in the various cases, wether it is 
(at least partially) due to intrinsic star-to-star physical differences or it can be fully accounted for by trivial effects.

   \begin{figure*}
   \label{SDSSphot}
   \includegraphics[width=6cm]{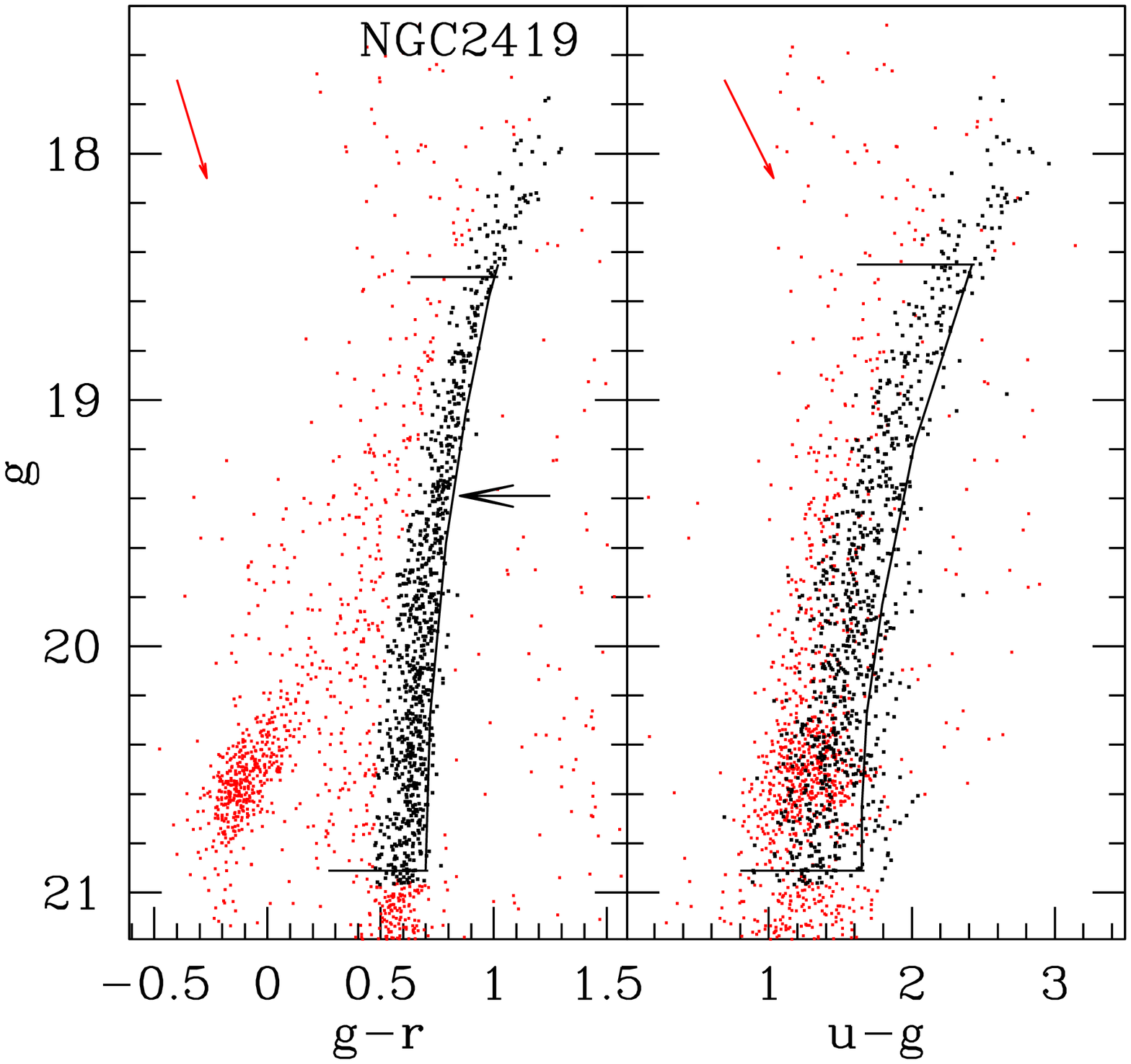}
   \includegraphics[width=6cm]{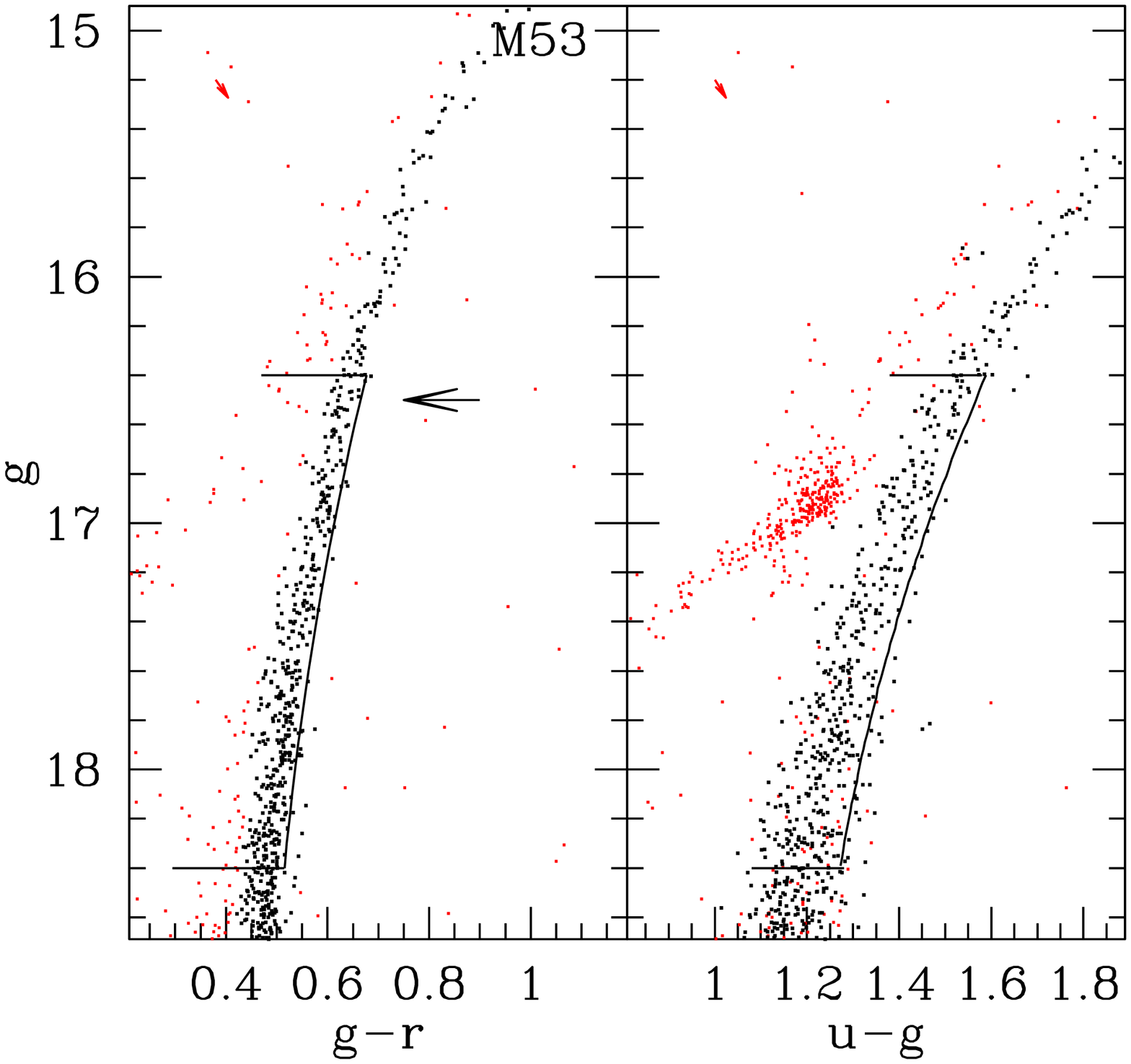}
   \includegraphics[width=6cm]{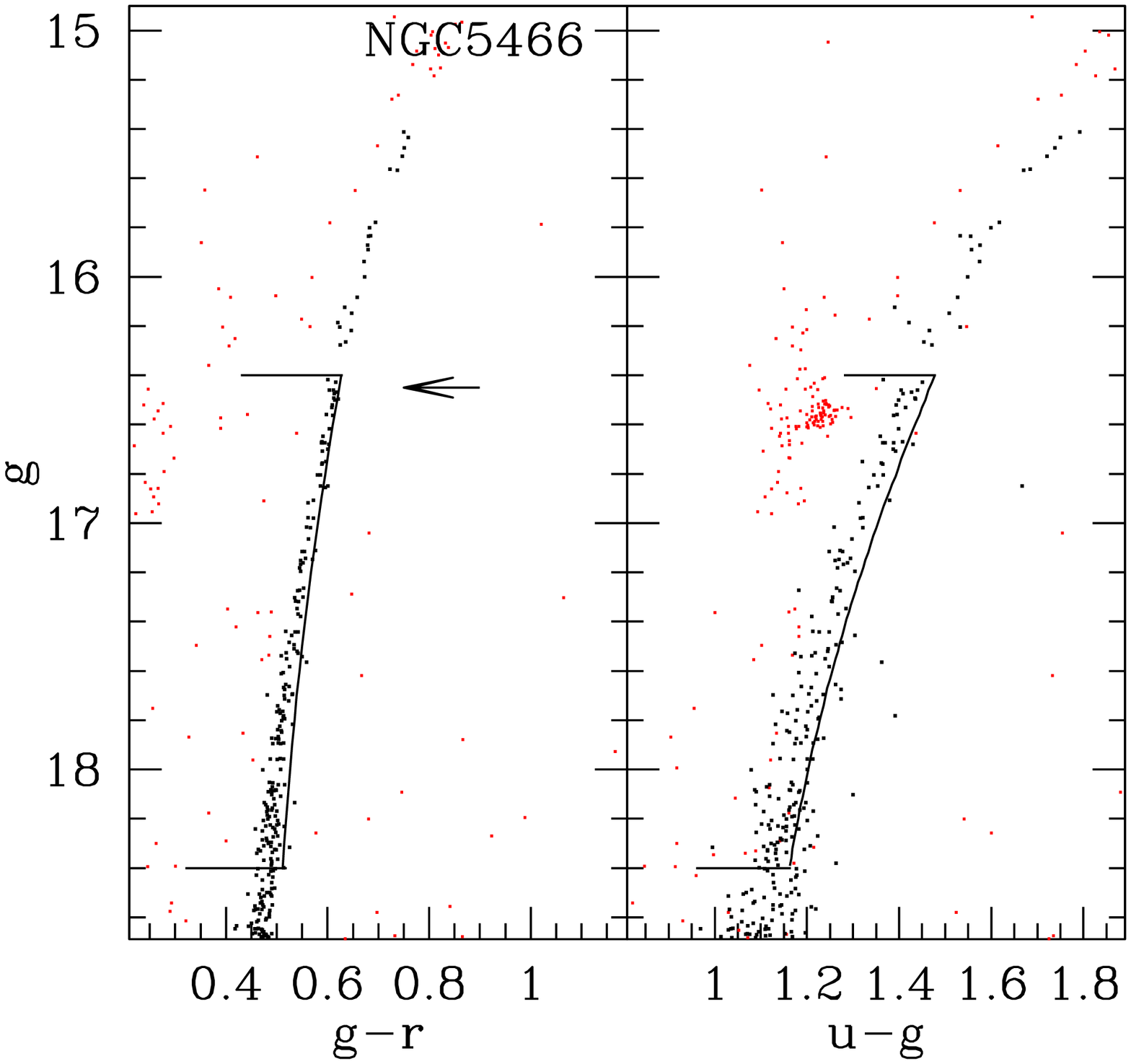}
   \includegraphics[width=6cm]{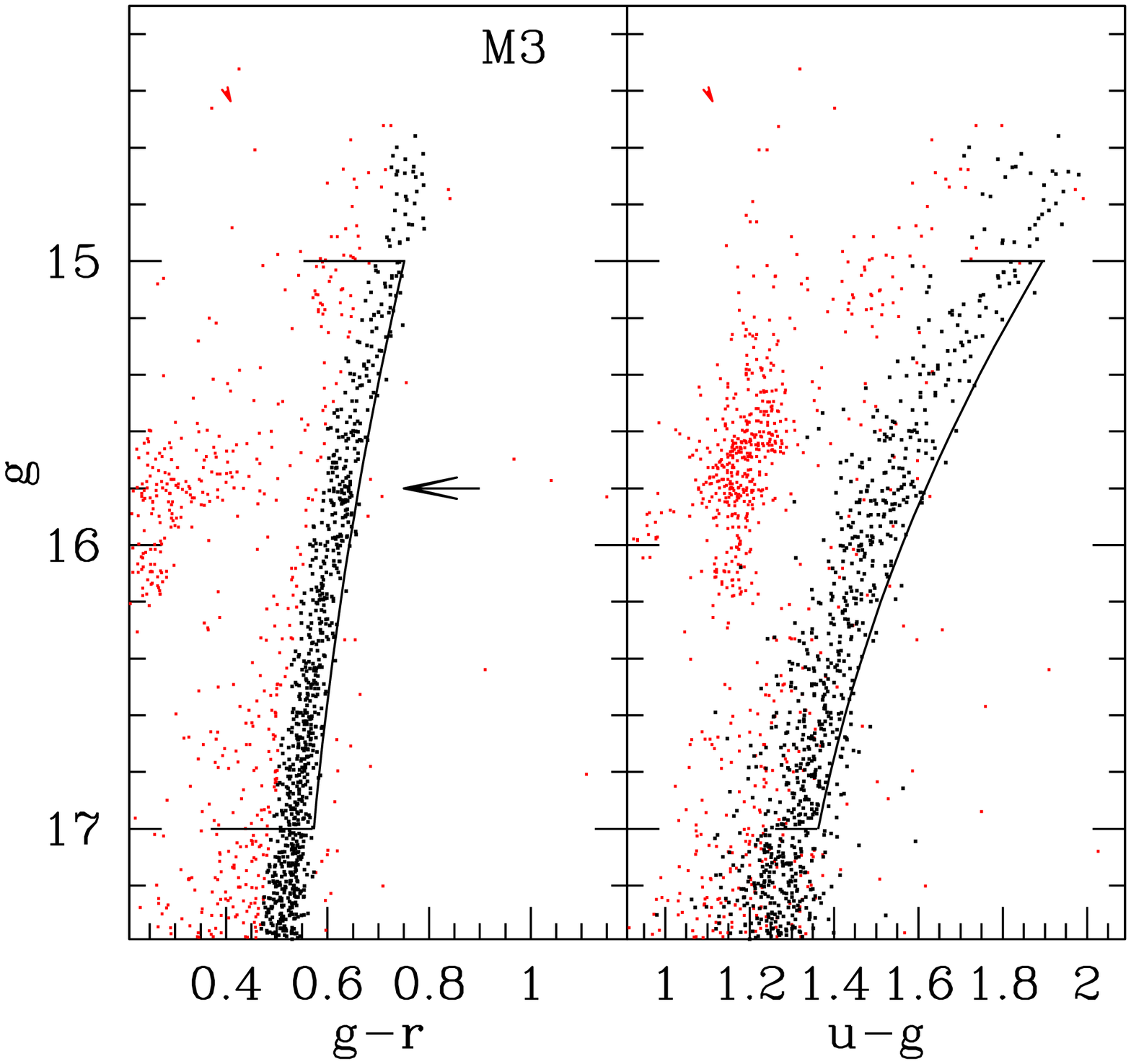}
   \includegraphics[width=6cm]{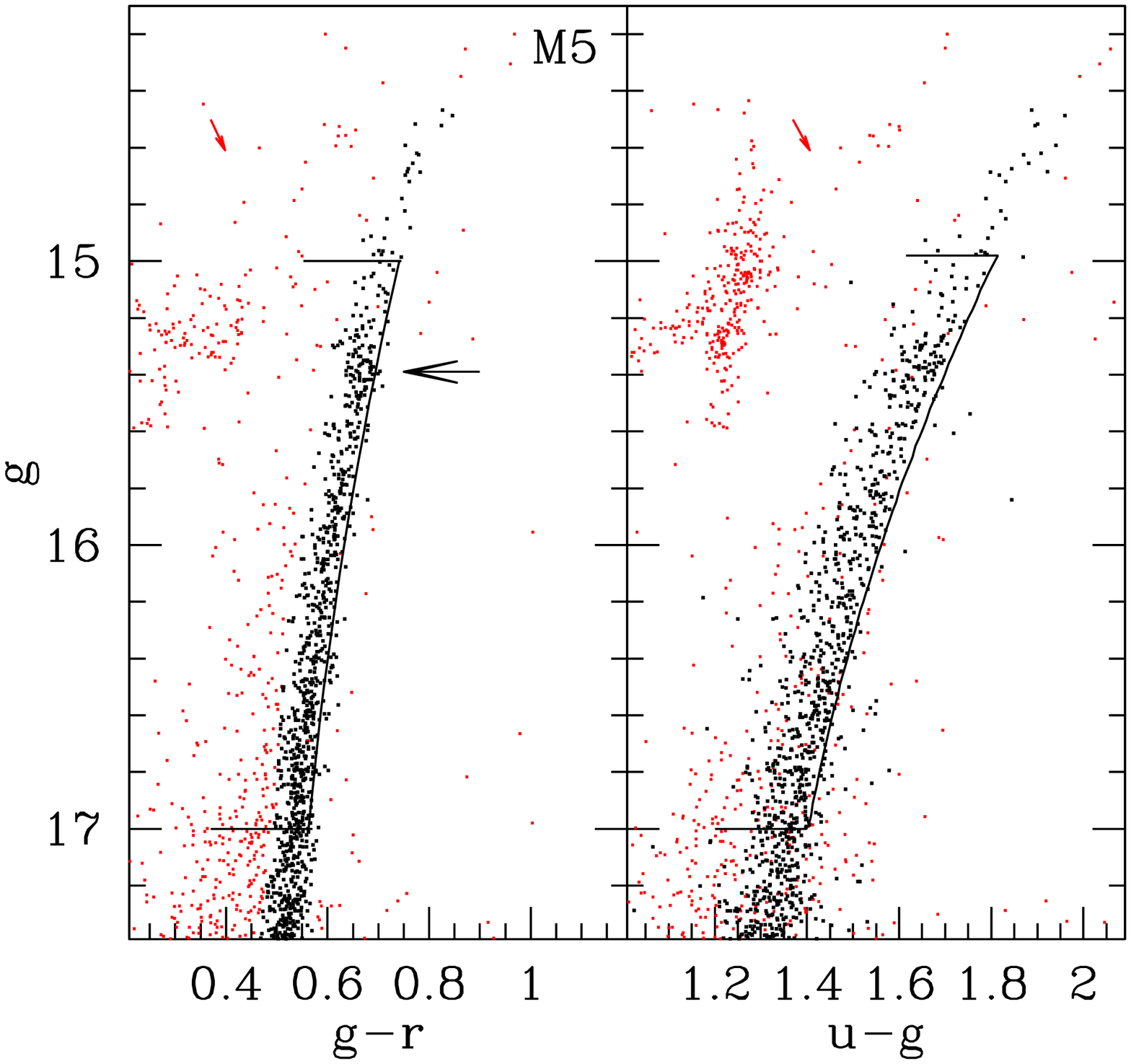}
   \includegraphics[width=6cm]{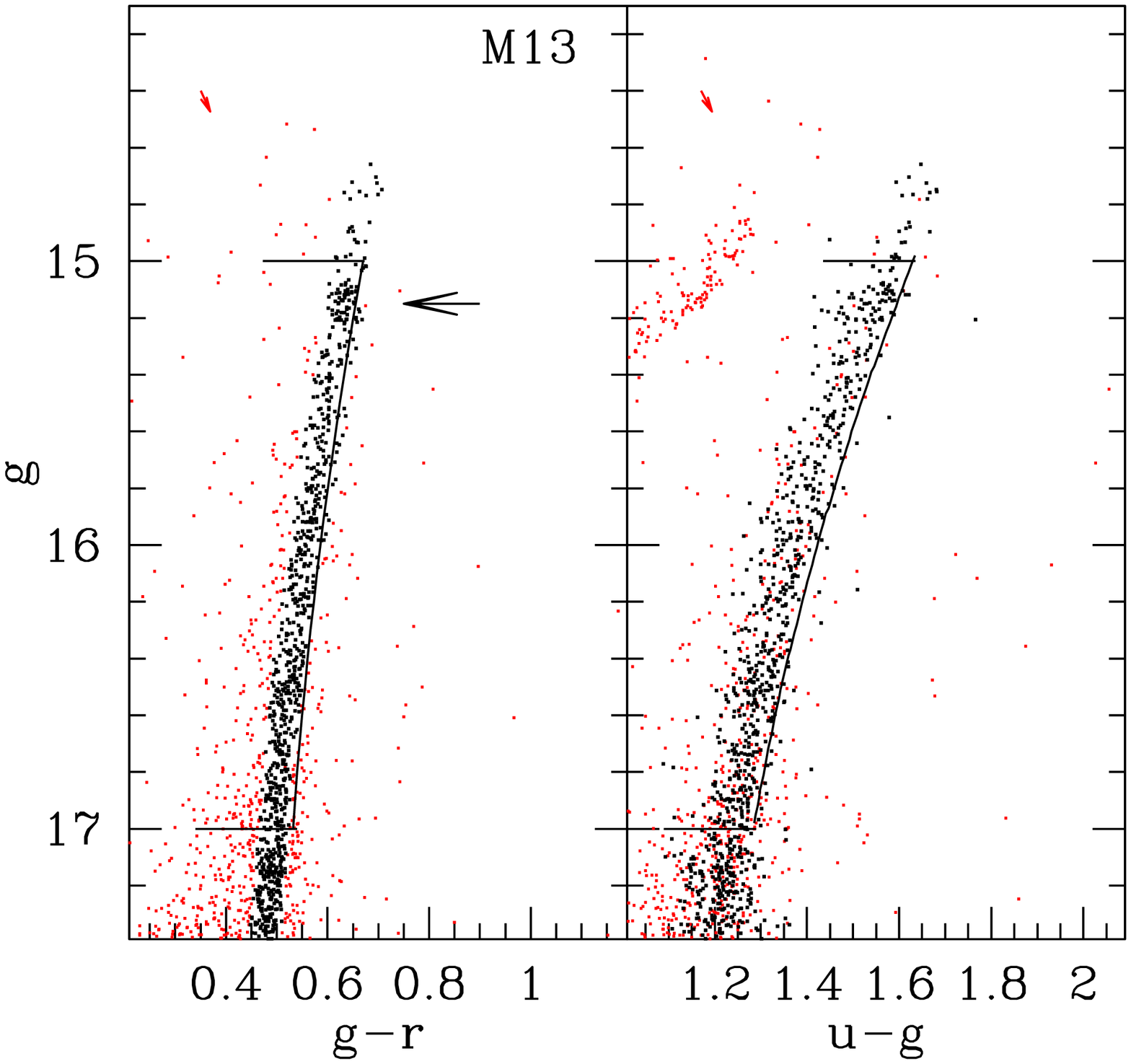} 
   \includegraphics[width=6cm]{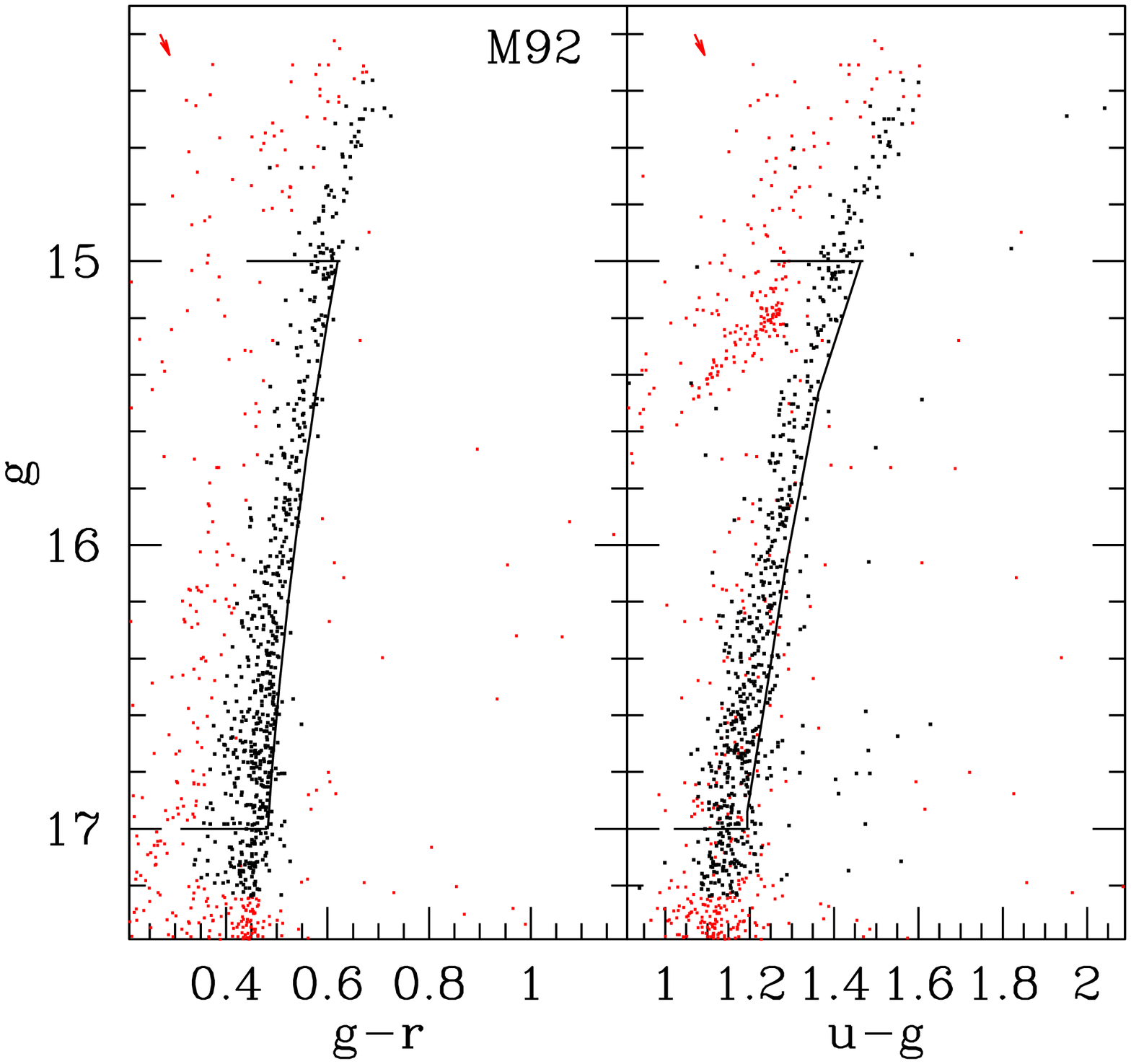} 
   \includegraphics[width=6cm]{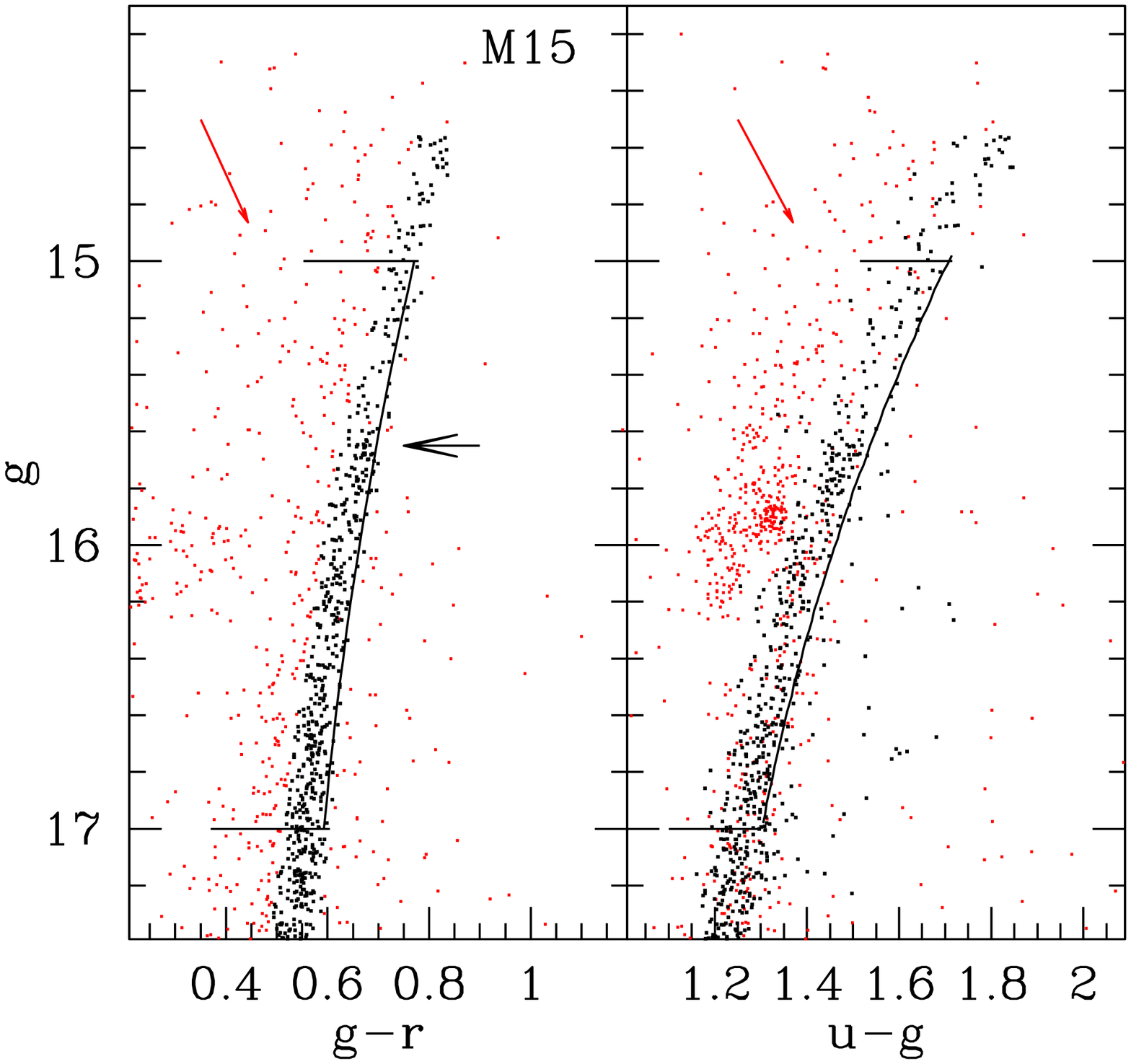}
   \includegraphics[width=6cm]{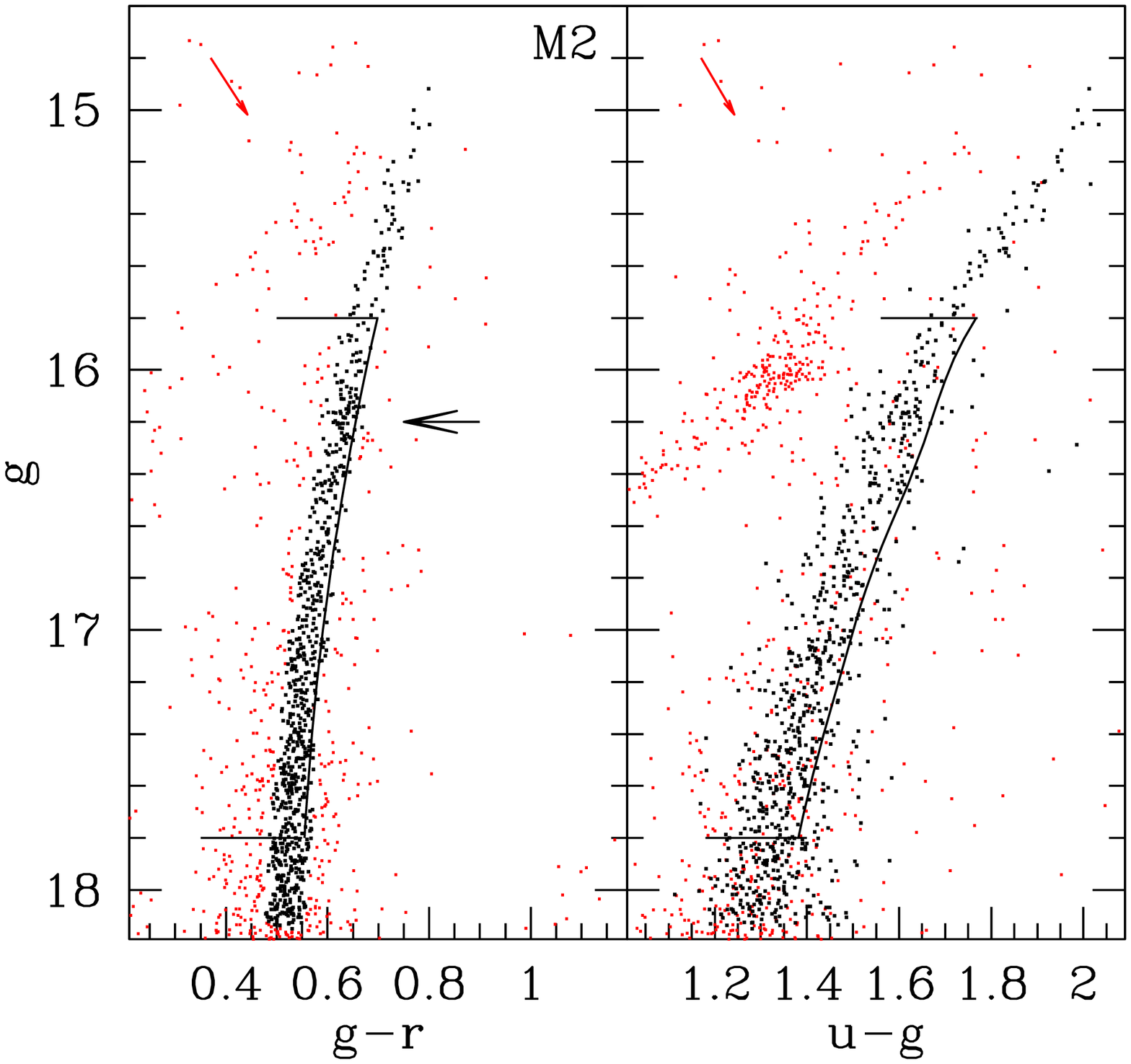}
   \caption{{\it g--r,g} and {\it g,u--g} CMDs for the GCs in our sample. In some cases, 
the RGB is truncated due to SDSS saturation. The red arrows in the upper right corner of each panel 
are the reddening vectors whose amplitudes correspond to the average E(B-V) values reported in Table~\ref{GCs}.
Horizontal black arrows mark the position of the RGB bump.
Stars selected as candidate RGB on the ({\it g,g--r}) CMDs are plotted as heavy dark dots, the remaining 
ones as lighter red dots. The curves approximately tracing the red edges of the RGBs are used 
as references to compute the color spread distributions shown in Fig.~\ref{Dist}; the the two 
horizontal segments display the portion of the RGB that is used to compute those distributions. In most 
cases, note that the asymptotic giant branch starts above the
bright end of the adopted selection box (NGC~2419 is an obvious exception).}           
    \end{figure*}

\section{Detecting anomalous u-g spreads}
\label{photo}

In Fig.~2, we present the {\it g,g-r} CMDs, flanked by their {\it g,u-g} counterparts for all 
the clusters listed in Tab.~\ref{GCs}. In all cases, two horizontal segments enclose the magnitude 
range to which we limit our analysis of the color spread: we tried to select similar portions of the RGB 
in all clusters,  possibly spanning a wide region below the horizontal branch. Given the different 
distances and reddenings of the various clusters, this was not always possible (in particular, for 
NGC~2419 which is far more distant than the other GCs in our sample). This gives rise to differences in the sensitivity 
of the adopted method as (a) the color spread corresponding to a given abundance spread is always observed 
to decrease with increasing luminosity along the RGB, virtually disappearing at the RGB tip 
\citep{yong08,marino2008,milone2010}, and (b) the same portion of the RGB occurs at different apparent
magnitudes in different clusters, which is indicative of different photometric accuracy. 

The line located approximately at the red edge of the RGB, within the two horizontal segments, 
is a ridge line following the curvature of the observed RGB, and is taken as a reference to 
compute the color spreads, that is the difference between the color of a given star and the 
color of the ridge line at the same magnitude, $\Delta_{col}$, where $col=g-r$ or $col=u-g$. 
Both $\Delta_{g-r}$ and $\Delta_{u-g}$ are computed only for stars selected as candidate 
RGB in the $g,g-r$ CMDs (heavier dots in Fig.~2). 
To limit the effects of possible spurious $u-g$ outliers,
we considered only stars for which $-0.2\le \Delta_{u-g}\le 0.05$; these limits were found to be 
appropriate for all clusters except NGC~2419 for which we adopted $-0.8\le \Delta_{u-g}\le 0.05$.

There are several cases in Fig.~2 in which a conspicuous broadening of the RGB 
in {\it u-g} with respect to {\it g-r} is apparent. However, as anticipated above, a few factors 
unrelated to physical differences among cluster stars may also (in principle) produce this 
effect. These factors and the methods we adopted to keep their effect under control can be 
summarized as follows:

\begin{enumerate}

\item {\em Field contamination.} The degree of contamination by Galactic fore/background stars 
is very modest, because of the combination between the relatively high (absolute) Galactic 
latitude of the considered clusters (all have $|b|>25\degr$ and four $|b|>70\degr$) and 
the relatively small area of the considered annular fields. 
We used the Galactic model TRILEGAL \citep{trilegal} to obtain 
a conservative estimate of the degree of contamination affecting 
the samples of candidate RGB stars considered in the present analysis (see Fig.~2). 
We found that the fraction of Galactic field stars in our samples is lower 
than 5\% for seven of the nine clusters, reaching 8\% for M~15 and 10\% for M~92.
Moreover, the effect of even this small degree of contamination 
should be minimized by the way in which we selected candidate 
RGB stars. In the following, we analyze the color spreads of candidate RGB stars selected 
in the {\it g,g-r} CMD as the most tightly clustered along the narrow RGB sequence in this plane 
(heavy dark dots in Fig.~2).
This means that we compare the {\it g-r} and {\it u-g} color spreads produced by  
{\em exactly  the same stars}. Hence, any field star artificially broadening the distribution of 
$\Delta_{col}$ in {\it u-g} should also have a similar effect on {\it g-r}.

\item {\em Blendings and artifacts.} Any photometry of crowded fields (especially ground-based 
seeing-limited ones) is expected to include some fraction of sources that 
are either blends of two (or more) fainter stars, or non-stellar sources, 
such as unresolved distant galaxies or flukes in the halo of bright stars,
both misclassified as stars.
While it is likely that our samples of RGB candidates include some of these sources, 
their effect is expected to broaden the distribution of $\Delta_{col}$ by a similar amount in both of the considered colors, given the adopted selection, as for the case of contamination described above.

\item {\em Differential reddening.} In principle, if there is a star-to-star difference in the 
degree of stellar extinction (due to spatial a variation in the reddening over the considered field 
of view) this will produce a larger color spread in {\it u-g} than in {\it g-r}, mimicking the effect we are 
looking for. However, the ratio of the expected spreads is quite small${\frac{\Delta E(u-g)}{\Delta E(g-r)}}=1.29$, 
meaning that a difference of 0.02 mag in E(B-V) would correspond to 0.019 in E(g-r) and 0.024 in E(u-g), 
i.e. a mere difference of 5 millimag between {\it g-r} and {\it u-g}.
The clusters considered here have, in general, low average reddening values (four having $E(B-V)\le 0.02$, 
seven $E(B-V)\le 0.06$, and all nine $E(B-V)\le 0.11$) and, as far as we know, no indication of differential 
reddening has ever been reported in the literature. In the cases for which we report the detection 
of significant {\it u-g} spread, the full width at half maximum (FWHM) of the distribution spread
in {\it u-g} is $\ga 2$ times larger than in {\it g-r}, 
more than the factor of 1.29 that can be attributable to differential reddening alone. Finally, 
the case of M~5 presented here as well as the previous cases reported in the literature 
(see Sect.~\ref{intro}) suggests that differential reddening does not play a major role in 
producing the UV-s, in the cases where this has been revealed up to now.
We conclude that the effects of differential reddening for the considered sample, should be negligible.

\item {\em Photometric errors}. Since all the SDSS observations are performed at fixed 
exposure time and RGB stars emit much less light in the near-UV than in the visible 
range\footnote{There are other factors concurring to lower the signal-to-noise ratio of $u$ 
observations with respect to $g$ or $r$ ones. For example: (1) CCDs are less sensitive in near-UV 
than in visible bands; (2) as in the SDSS the photometry of a given field is taken simultaneously 
in all the $ugriz$ passbands and the seeing worsens at shorter wavelengths, $u$ images hence experiencing the 
worst seeing; (3) at fixed atmospheric transparency conditions and air mass, $u$ light suffers from the 
highest amount of atmospheric extinction.}, any 
given star in our sample has larger photometric errors in {\it u} than in either {\it g} or {\it r}: as a consequence, 
the color spread due to photometric errors is larger in {\it u-g} than in {\it g-r}. We see below that 
this is the most serious problem in the present analysis, as there are cases in which a 
large observed UV-s can be fully accounted for by (relatively) large photometric 
errors that hide any (possible) underlying signal associated to real differences among stars. 
To take this effect into due 
account, we divided the color spread of each star by the associated photometric uncertainty in each 
color $\Delta^{'}_{col}= {\frac{\Delta_{col}}{\sigma_{col}}}$ ({\em normalized color spread}), 
which should be (approximately) expressed in units of standard deviations. 
Comparing $\Delta^{'}_{col}$ distributions in {\it g-r} and {\it u-g}, we can check whether there is any significant 
UV-s in addition to that due to photometric errors; if the latter were the main contributor 
to the observed {\it u-g} spread, the two distributions should be indistinguishable.

\end{enumerate}

\begin{figure*}
   \centering
   \includegraphics[width=\columnwidth]{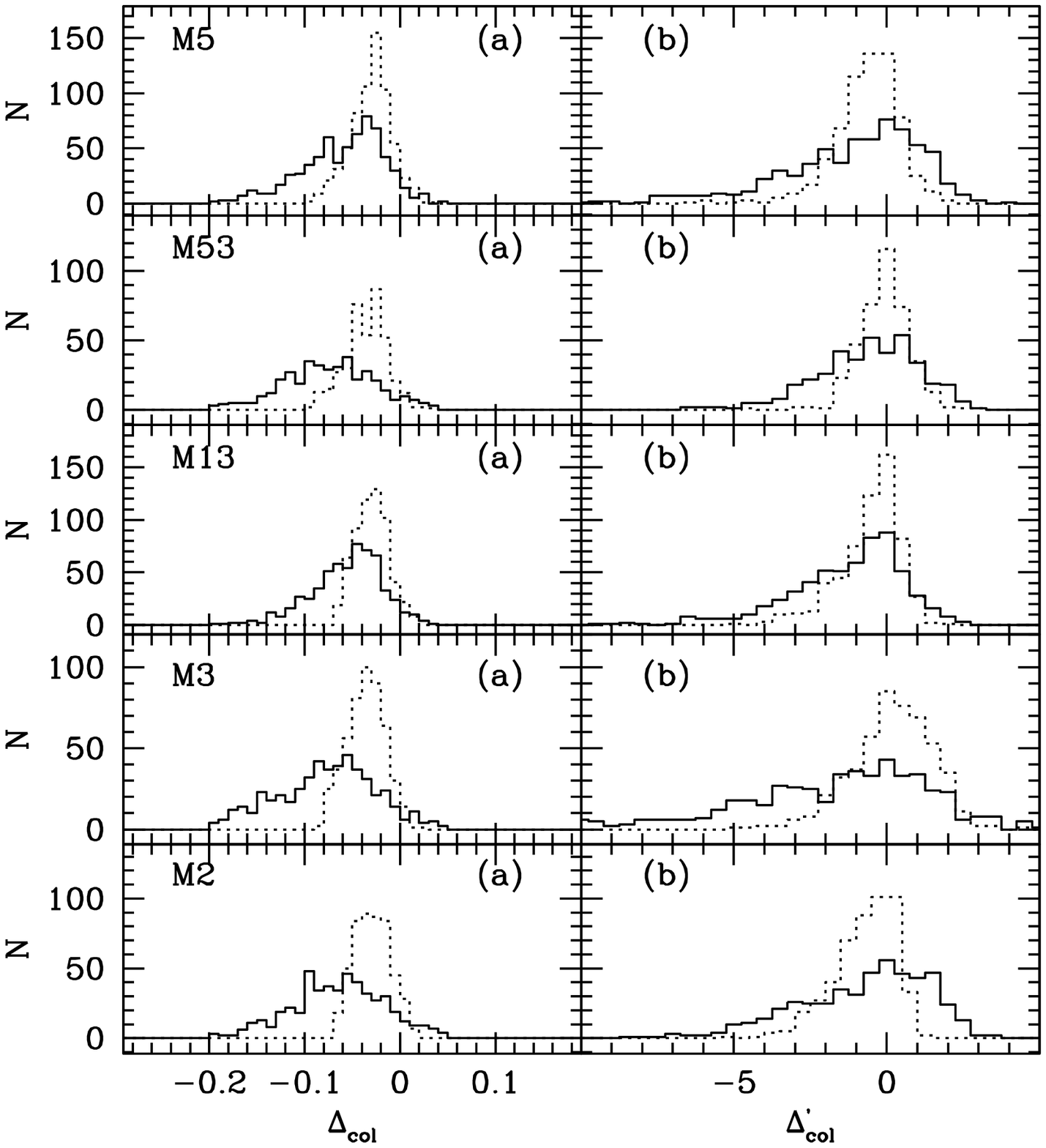}
\includegraphics[width=\columnwidth]{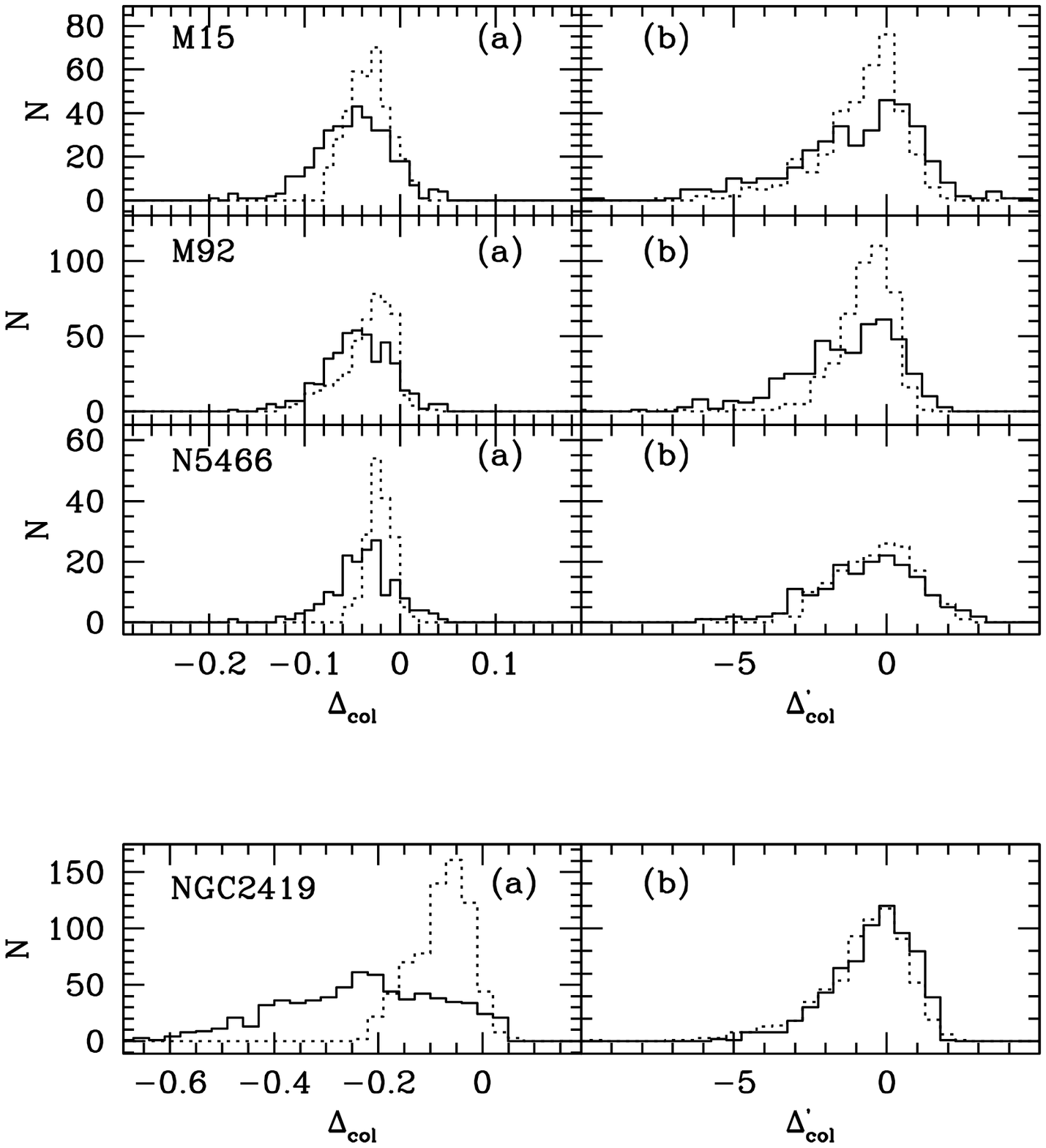}
    \caption{Distributions of $u-g$ (solid histograms) and $g-r$ (dashed histograms) color spreads 
(with respect to the RGB fiducials of Fig.~2) 
for the clusters in our sample. Panels marked with (a) 
show the distributions of absolute color spreads, while those marked with (b) show the {\em normalized} color spreads, 
in units of $\sigma$.}
         \label{Dist}
   \end{figure*}

In Fig.~\ref{Dist}, we plot the distribution of $\Delta_{col}$ and $\Delta^{'}_{col}$ for all the 
considered clusters. Histograms in {\it g-r} are plotted as dotted lines, and those in {\it u-g} as continuous 
lines. The $\Delta^{'}_{col}$ distributions have been shifted by (small) arbitrary amounts to ensure that 
their maxima approximately coincide with $\Delta^{'}_{col}=0$, to allow a more direct 
comparison between the distributions in the two color indices\footnote{In particular, the {\it g-r} 
distribution has been shifted to ensure that its well defined maximum occurs at $\Delta^{'}_{col}=0$, 
while for the broader {\it u-g} distributions, with less clear peaks, we searched for a compromise 
between placing the peak at $\Delta^{'}_{col}=0$ and matching the right (red) edge of the two 
distributions, to make the comparison easier.}. We shall see below that the significance of the 
detected differences is maintained independently of the adopted shifts.

It may be useful to start the discussion of Fig.~\ref{Dist} from the pair of panels in the 
lower right corner (NGC~2419). As readily visible in the CMD of Fig.~2, there 
is an impressive broadening of the RGB color spread passing from {\it g-r} to {\it u-g} in this case.  The broadening 
in {\it u-g} color is so large that we had to adopt a different horizontal scale in this panel to accommodate the bulk of the 
$\Delta_{u-g}$ distribution.
However, the distributions of $\Delta^{'}_{col}$ are indistinguishable, with essentially all the stars 
lying within $\pm 3\sigma$ of the mean, as expected for (approximately) normal errors\footnote{we
note that this is true for all the {\it g-r} distributions of $\Delta^{'}_{col}$.}. 
Hence, all the UV-s observed in NGC~2419 can be accounted for by the effect of photometric errors. 
We note that {\em this does not mean that there is no underlying spread in light-element abundance, 
but that the photometric accuracy is not sufficient to reveal it, if it exists}. The same is true for NGC~5466, while the case of 
M~15 is discussed in more detail below.

On the other hand, in all the cases considered in the left raster of panels pairs of Fig.~\ref{Dist} 
(M~5, M~53, M~13, M~3, and M~2), plus M~92 in the right raster of panels, the differences in the $\Delta_{col}$ 
distributions correspond to 
significant differences in the distributions of $\Delta^{'}_{col}$, implying that the detected UV-s 
cannot be entirely accounted for by observational effects, thus requiring the presence of 
some physical star-to-star difference. 
In particular, the $\Delta^{'}_{u-g}$ distributions display extended tails toward the blue 
that are completely lacking in the remarkably symmetric g-r distributions. In the case of M~5, we 
know from Sect.~\ref{data} that the observed {\it u-g} spread correlates with Na abundances \citep[and it is 
likely to be caused by differences in N abundances,][]{yong08}.
It is interesting to note that the $\Delta_{col}$ distribution of this cluster appears to be bimodal 
\citep[as in the case of M~4][]{marino2008}, even if it is not possible to firmly establish 
the statistical significance of this feature. Hints of multi-modality are indeed visible in all the 
$\Delta_{col}$ (and $\Delta^{'}_{col}$) distributions of these six clusters.
 

\begin{table*}
\caption{Dimensions of the samples and results of KS tests.}             
\label{sigma}      
\centering          
\begin{tabular}{l c c c c c c }
\hline \hline      
   
NGC 		&   $N_{stars}$	 & Shift     	& $P_{KS}^{phot}$     		& $N_{UV-blue}$ &  $N_{UV-red}$ &$P_{KS}^{rad}$   \\
\hline
   M~5  	 &  652		&  0.2	 	& $1.1 \times 10^{-11}$	        &199		&453		&$2.0 \times 10^{-6}$ \\ 
   M~2  	 &  492		&  --0.3 	& $6.0 \times 10^{-10}$ 	&132		&360		&$<1.0\times 10^{-11}$\\  
   M~3  	 &  501  	&  1.5  	& $3.9 \times 10^{-10}$ 	&206		&295		&$3.8 \times 10^{-10}$\\  
   M~13 	 &  596		&  0.4  	& $5.7 \times 10^{-8}$ 		&185		&411		&$1.3 \times 10^{-4}$\\  
   M~92   &  442		&  0.4  	& $3.8 \times 10^{-7}$		&139		&303		&$<1.0 \times 10^{-11}$ \\    
   M~53 	 &  394  	&  0.4  	& $2.9 \times 10^{-7}$ 		&75		&319		&$5.7 \times 10^{-11}$\\  
   M~15	 &  371		&  0	 	& $8.9 \times 10^{-4}$		&110		&261		&$7.8 \times 10^{-11}$\\    
   NGC~5466  &  172	&  0.2  	& 0.60 				&37		&135		& 0.15 \\ 
   NGC~2419  &  693	&  --0.2  	& 0.79 			        &94		&599		& 0.24\\ 

\hline                  
\end{tabular}
\tablefoot{$N_{stars}$: total number of candidate RGB stars selected for the analysis of color spread. 
''Shift'' is the differential shift in $\Delta^{'}_{col}$ that is found to maximize 
$P_{KS}^{phot}$. $N_{UV-blue}$ and $N_{UV-red}$ are the number of stars having $\Delta^{'}_{col}<2.0$ 
and  $\Delta^{'}_{col}\ge 2.0$, respectively.
$P_{KS}^{phot}$ and $P_{KS}^{rad}$ are defined in the text. Clusters are listed in order of increasing $P_{KS}^{phot}$.}
\end{table*} 


A Kolmogorov-Smirnov (KS) test would be the most straightforward non-parametric way to 
quantitatively establish the statistical significance of the detected differences between 
the {\it u-g} and {\it g-r} $\Delta^{'}_{col}$ distributions. However, it is well known that this test 
is very sensitive to offsets between distributions and the shifts adopted in Fig.~\ref{Dist} 
are quite arbitrary. In principle, a given choice of the shift between the two 
distribution may add spurious significance to their mutual difference measured by the KS 
test. To circumvent this problem, we proceeded as follows: (a) keeping the {\it g-r} distributions 
shown in Fig.~\ref{Dist} fixed, we moved the {\it u-g} distribution from $-5$ to $+5$ $\Delta^{'}_{col}$ units 
in steps of 0.1; (b) at each step, we performed the KS test computing the probability that the two 
samples are drawn from the same parent population, $P_{KS}^{phot}$; and, (c) we adopted the value of 
the shift that {\em maximize} $P_{KS}^{phot}$, i.e. the shift that {\em minimize} the 
significance of the difference between the two distributions. In this way, we are guaranteed 
that the residual differences considered by the KS test are genuine differences {\em in the 
shape of the distributions}, not due to unphysical shifts. The adopted differential shifts 
and the corresponding values of $P_{KS}^{phot}$ are reported in Tab.~\ref{sigma}. 
The significance of the difference in the {\it u-g} and {\it g-r} distributions of $\Delta^{'}_{col}$ 
is very high for all the clusters whose distributions are plotted in the left 
hand of Fig.~\ref{Dist} plus M~92 ($P_{KS}^{phot}<10^{-4}$).
In particular, we note that the difference is also significant 
in the case of M~15: the probability that the {\it u-g} and {\it g-r} distributions of 
$\Delta^{'}_{col}$ 
measured in this cluster are drawn from the same parent population is just 0.08\%. 
A careful inspection of the $\Delta^{'}_{col}$ distributions for this cluster in Fig.~\ref{Dist} reveals that while 
the wings  virtually coincide, the core of the {\it g-r} distribution 
is far more peaked than its {\it u-g} counterpart, which exhibits hints of bi-modality (compare 
with the cases of NGC~5466 and NGC~2419, where the distributions nearly coincide in both 
the wings {\em and} the core). Hence, while the amplitude of the effect is insufficient 
to provide an obvious color segregation in the CMD of this cluster (see Fig.~\ref{Spectro}), 
the underlying signal is there and can be revealed once the much larger photometric sample is considered 
and the observational effects are properly taken into account, in agreement with the results of \citet{yong08}. 
This conclusion is strongly 
supported by the behavior of the radial distributions of RGB stars as a function of their color spread in 
this cluster, as discussed in the following section.

We note that the four clusters displaying the most obvious and significant difference in their 
$\Delta^{'}_{u-g}$ and $\Delta^{'}_{g-r}$ 
distributions have metallicity around [Fe/H]$=-1.5$, 
while M~53, M~92, and M~15, which have larger $P_{KS}^{phot}$,
are significantly more metal-poor ([Fe/H]$\la -2.0$). 
This is likely due to the weakening of the intrinsic UV-s effect at low metallicity, discussed in Sect.~\ref{data-spec}.

\subsection{The radial distribution of UV-red and UV-blue stars}
\label{radsec}

As anticipated in Sect.~\ref{intro}, it would be very interesting to check whether RGB stars 
with different $u-g$ colors (at the same magnitude) may have different radial distributions. According to some
models of the early enrichment of GCs, at the end of the enrichment 
phase (lasting a few hundred Myr), stars from the first generation (P) 
should be significantly less concentrated toward the center of the cluster 
than stars of  later generation(s) (I+E), born from material 
polluted by the ashes of P stars \citep{dercole08}. While it may be expected that this 
difference should have been largely erased long ago by the
dynamical evolution of the cluster \citep{dercole08,decressin08}, the (sparse) 
available observational evidence about radial distributions of the two 
populations is in qualitative agreement with the prediction of the above 
quoted model. For example, \citet{kravtsov10a,kravtsov10b} found that the 
stars on the red side of of the broad U-I distribution of the RGB of NGC~1261 
and NGC~3201  are significantly more centrally concentrated than those on the blue side. In the case of NGC~3201,
\citet{euge32} used detailed Na, O abundances from high resolution spectroscopy to show that Na-rich RGB
 stars are redder (in U-I) and more centrally concentrated than Na-poor RGB stars.

\citet{car09b} merged the spectroscopic sample of RGB stars in
19 GCs, normalizing the distance of each star from the center of its cluster 
by the cluster $r_h$, and compared the radial distributions of P, I, 
and E stars. They found that I stars are significantly more centrally 
concentrated that P stars, while E stars appear slightly {\em less} concentrated than 
P stars. However, as discussed in detail by \citet{car09b}, this result may be affected by serious selection biases, inherent to the process of efficient fiber allocation in multi-object spectroscopy of Galactic GC stars. 

A difference in the radial distribution of the two sub giant branches of NGC~1851
\citep[whose stars are presumed to differ in terms of CNO abundance;][]{santi} was 
detected by \citet{zoccali09}, but the result was not fully 
confirmed by \citet{milone09b}.
However, \citet{eu1851} find that the radial distribution of the RGB stars they analyzed depends on 
their iron abundance, more metal-poor stars being more centrally concentrated than their more metal-rich counterparts.

Different populations with different radial distributions are known to be present in 
$\omega$~Centauri \citep{pancino03,sollom,villanova}, but this system is more complex and may have a different 
origin with from classical GCs \citep[see][and references therein]{m54}.

   \begin{figure}
   \centering
   \includegraphics[width=\columnwidth]{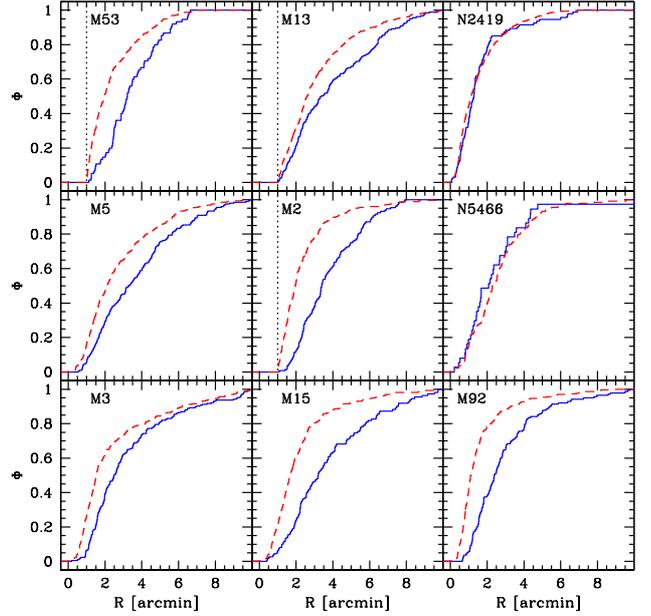}
    \caption{Comparison between the radial distribution of UV-blue (continuous blue line) 
and UV-red (dashed red line) for all the considered clusters. The data within the radius marked by the dotted lines 
in the plots of the distributions for M~53, M~2, and M~13 were not included in the analysis (see Sects.~\ref{data} and \ref{data-spec}). 
See Table~\ref{sigma} for the 
significance of the detected differences, as measured by the KS test.}
         \label{rad}
   \end{figure}

To follow-up this line of investigation with our sample, we divided the selected RGB 
candidates in each of  the considered clusters into two sub-samples, according to the 
value of their  normalized {\it u-g} spread 
$\Delta^{'}_{col}$. For brevity, we dub UV-blue the stars having  $\Delta^{'}_{col}<-2.0$ 
and UV-red those having $\Delta^{'}_{col}\ge -2.0$, in the scale of Fig.~\ref{Dist} (i.e. using 
the shifts adopted there). It is important to recall that any observational effect potentially 
affecting the radial distributions (notably, the radial variation in the degree of completeness, 
due to the increase in crowding toward the center) must affect the two sub-samples exactly in the 
same way, as they have the same distribution of magnitudes.

The comparison between the radial distributions of UV-blue and UV-red RGB stars for all the 
considered clusters is presented in Fig.~\ref{rad}; the probability that the two distributions 
are drawn from the same parent population according to a KS test ($P_{KS}^{rad}$) is reported 
in the last column of Tab.~\ref{sigma}. In all the clusters for which we detected a significant intrinsic {\it u-g} 
spread (including M~15), the UV-red population is obviously more centrally 
concentrated than the UV-blue one, with $P_{KS}^{rad}$ always lower (and in most cases {\em much} 
lower) than 0.02\%. We note that this result is very insensitive to the actual choice of the 
$\Delta^{'}_{col}$ 
threshold; the difference remains highly significant for a large range of adopted thresholds 
(in particular, for the whole range $-3.0\le\Delta^{'}_{col} \le0.0$).

   \begin{figure}
   \centering
   \includegraphics[width=\columnwidth]{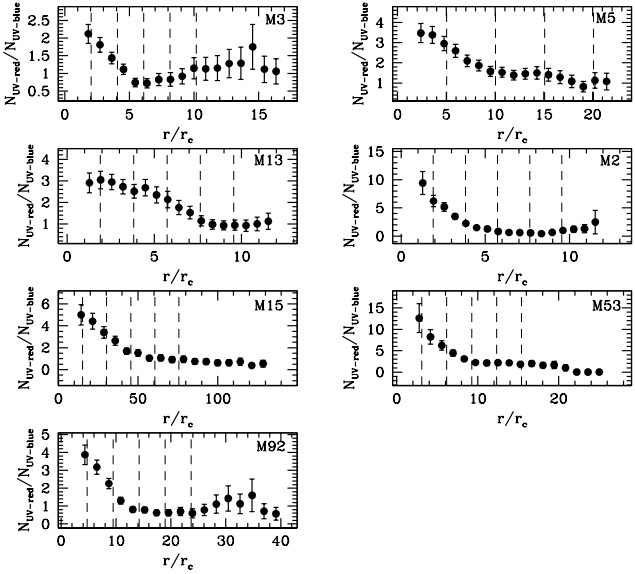}
    \caption{Ratio of the number of UV-red to UV-blue stars as a function of distance from the cluster center, 
for the seven clusters in which we detected a significant intrinsic UV-s. The radial coordinate is expressed in units 
of cluster core radii, the dashed vertical lines marks the radial distances corresponding to 1, 2, 3, 4, 5 
half-light radii \citep[$r_c$ and $r_h$ from][]{harris1996}. The ratio is computed in 
radial bins $2.5\arcmin$ wide in steps of $0.5\arcmin$. Note that the actual value of the ratio 
depends on the adopted threshold between UV-red and UV-blue stars: here we adopted 
$\Delta^{'}_{col}=-2.0$ as in Tab.~\ref{sigma} and Fig.~\ref{rad}.     }
         \label{ratio}
   \end{figure}

This result clearly provides further support to the physical significance of the 
{\it u-g} spread: 
it is very hard to conceive how any spurious observational effect can be associated with such a strong 
difference in the radial distribution. Moreover, it suggests that the higher degree of central 
concentration of UV-red (putative I+E) stars with respect to UV-blue (putative P) stars may be a 
general characteristic of all GCs where intrinsic
UV-s can be detected, 
therefore any model intended to explain the origin of the spread in light-element abundances in 
GCs should also be able to reproduce this feature.

Fig.~\ref{rad} aims to demonstrate the high level of statistical significance of the detected differences between
the radial distributions of UV-red and UV-blue stars. To allow a more direct comparison with the predictions 
of chemo-hydro-dynamical models \citep{dercole08,decressin08,dec2}, in Fig.~\ref{ratio} we present the radial 
profile of the ratio of the number of UV-red to UV-blue stars 
($\frac{N_{UV-red}}{N_{UV-blue}}$), where the radial coordinate is expressed in 
units of the clusters core radii ($r_c$), and the distances of 1, 2, 3, 4, and 5 half-light 
radii ($r_h$) are also indicated (vertical dashed lines). This ratio should approximately 
scale as the ratio of second generation(s) (I+E) to first generation (P) stars, whose radial distribution 
is a typical outcome of the considered models. 
However, we stress 
that while the overall shape of the observed profiles does not change much when different 
thresholds between UV-blue and UV-red stars are 
adopted\footnote{Over the range 
$-3.0\le \Delta^{'}_{col}\le -1.0$. In particular, the slope of the inner 
rising branch of the profile changes with the adopted threshold (the relative 
height of the central peak grows by a factor of $\la 3-4$ changing the threshold from -1 to -3) 
but the radius where the profile flattens remains unchanged.
In Fig.~\ref{ratio}, we have adopted $\Delta^{'}_{col}=-2$, as above.}, the true 
values of $N_{UV-red}/N_{UV-blue}$ depends on the adopted threshold, hence they cannot 
be directly compared with second-to-first generation ratios computed elsewhere. The only
safe conclusion that we can draw here is that UV-red stars are more abundant
than UV-blue stars (in the innermost $\sim 3-4~r_h$) for any threshold  $\Delta^{'}_{col}\le -1.0$, 
albeit with
cluster-to-cluster differences of a factor of a few; this is in qualitative agreement 
with the predictions by \citet{dercole08} and \citet{dec2}, and with the observations by \citet{car10}. 

The profiles shown in Fig.~\ref{ratio} cover the range given by $\simeq 0.5-1~r_h$ and $> 4~r_h$. In nearly all cases, 
the profiles display a relatively steep decline from the innermost bin out to a radius of $\sim 3~r_h$ where they flatten 
at a level $N_{UV-red}/N_{UV-blue}\la 1$, remaining approximately flat out to the last observed point. Hints of another increase 
at $r_h\ga 10~r_h$ are seen in the profiles of M~3 and M~2 but their significance seems only marginal, if any. A notable exception
 to this general trend is provided by M~13, whose profile is nearly flat out to $\sim 3~r_h$, then declines into another 
flat branch at $r\sim 4~r_h$. 
 
It is interesting to compare the observed profiles with the model predictions shown
in Fig.~18 of \citet{dercole08}. In that specific model, after 25 half-light relaxation times of evolution, 
the profile of the second-to-first generation number ratio is nearly flat within $r\simeq 0.5~r_h$ (a region 
always entirely enclosed in the innermost bin of our profiles, or not even included in the considered sample, 
in the cases of M~2 and M~53), then has declined by a factor of a few by $\simeq 2~r_h$, 
at the limit of that figure. This is in fair agreement with the observed profiles\footnote{However, it is expected 
that the inner flat region of the profile would progressively extend as dynamical evolution proceeds beyond 
the $25~t_{rh}$ time lapse considered in that simulation (E. Vesperini, private communication).}; a broad 
agreement is also found with some of the models presented in \citet[][see the middle bottom and right 
bottom panels of their Fig.~1]{dec2}. A detailed comparison with models 
is far beyond the scope of the present paper. On the other hand, we feel that Fig.~\ref{ratio} provides a 
very useful set of observational constraints that must be reproduced by models of GC formation. 
Unfortunately, our data do not sufficiently probe the innermost regions ($r\le 0.5~r_h$) of the considered 
clusters; and complementary HST observations are probably needed to check this part of the profiles. Finally, 
a fully meaningful comparison with the profiles of Fig.~\ref{ratio} would require dedicated models for 
each cluster, taking into account its specific structural and dynamical properties, as well as accounting 
for its evolution for a Hubble time.

\subsection{Comments on individual clusters}
\label{individ}

In this section, we briefly report on previous results about abundance and/or UV color spreads in the clusters considered in the present paper. 

Five of our clusters were also considered in the study by \citet{yong08}, namely M~5, M~3, M~13, M~15, and M~92.
These clusters display a 
significant spread at all evolutionary stages in the {\it V, $c_{y}$} CMD  \citep[from photometry by][]{grundahl99} . According to \citet{yong08}, the observed spreads are comparable to that seen in
NGC~6752, suggesting that all these clusters exhibit a $\simeq$2.0 dex dispersion in [N/Fe], i.e., the value found in NGC~6752.

\citet{grundahl98} provided evidence of stars from two different populations in the horizontal branch of M~13.
In particular, they noted that at any given $(b-y)_{0}$ color, there is a large spread in the $c_{0}$ index 
(defined as $c_{0}=c_{1}-0.2E(b-y)$) that they interpreted as being due to star-to-star-variations in CNO abundances.\\

Turning to spectroscopic analyses, \citet{martell08} showed that M~53 
has a broad but not strongly bimodal distribution of CN band strength,
with CN and CH band strengths that are anticorrelated for unevolved stars.\\

For the well-studied cluster M~5, \citet{smith83} showed that the cyanogen distribution is strongly 
bimodal for a sample of 29 stars near the RGB tip.
The classic Na/O anti-correlation has been 
found by \citet{sneden92} and confirmed by \citet{ivans01}. 
This was further correlated with the 
CN strength index, $\delta$S(3839): stars with larger CN indices also have larger 
Na and Al abundances and lower O abundances than stars with lower CN indices.
Finally, \citet{briley92} and \citet{cohen02} found strong (and anti-correlated) variations in the abundances of 
C and N of stars down to the base of the RGB of M~5.\\

An extended Na/O anti-correlation was found also in the extremely metal-poor cluster 
M~92 \citep{sneden91}, as well as strong (and anticorrelated) variations 
in the abundances of  C and N \citep{carbon82}.\\

Spectroscopic observations revealed 
star-to-star variations in the abundances of the CNO group 
elements among the M~3 giants \citep{pilachowski01}, with both oxygen-rich and oxygen-poor stars coexisting 
in the cluster \citep{kraft92,cohen05}.\\
\citet{lee00} claimed the 
existence of a C-N anticorrelation among stars on the lower RGB of M~15, although 
no bimodality is found \citep{cohen05}.
\citet{kayser08} showed a weak indication of a bimodality in the CH-CN plane 
(two clumps separated at CN$\sim$--0.6), although the 
observational errors are large compared 
to the separation of the two clumps.
Pancino et al. (2010, submitted to A\&A) detected a clear and 
bimodal CN and CH anticorrelation among the unevolved stars measured by 
\citet{kayser08}.
Finally, \citet{smith90} found a CN-CH band strength anticorrelation and a 
possible bi-modality in their sample of red giants in the cluster M~2.\\

We note that for all the clusters in which we detected a significant 
intrinsic {\it u-g} spread, there was previous evidence of inhomogeneity in their stellar population reported in literature, 
supporting our findings.

Previous detections of differences in radial distributions between different populations in GCs were limited to 
the cases of $\omega$~Cen, NGC~1851, NGC~1261, and NGC~3201, as discussed in Sect.~\ref{radsec}. 
The results presented here more than doubles the number of GCs where these differences are detected at a very 
high degree of significance.

 \section{Summary \& conclusions}\label{conclusion}

We have used publicly available {\it u,g,r} photometry from \citet{an2008} to search
for anomalous spread in near UV color ({\it u-g}) along the RGB of nine
high-latitude, low-reddening, and well populated Galactic GCs. This anomalous
spread (UV-s) was detected before in some clusters, using colors including  other broad/intermediate 
band filters, such as Johnson's U or Str\"omgren {\it u}, and it was shown to be
associated with the well known spread in the abundance of light elements
\citep[C, N, O, Na, etc.; see][and references therein]{car09b}. 
The main results
of our analysis can be summarized as follows:

\begin{enumerate}

\item We have introduced a method to remove the effect of photometric errors from the
observed color spread, normalizing the color residual about a fiducial RGB line 
of each considered star by the associated photometric error 
($\Delta^{'}_{col}$). By comparing the distribution of $\Delta^{'}_{u-g}$ with
that of  $\Delta^{'}_{g-r}$ of the same sample of candidate RGB stars, we
revealed anomalous {\it u-g} spreads in seven of the nine clusters. The observed
effect has a very high statistical significance in all seven cases: according to
Kolmogorv-Smirnov tests, the probability that the observed $\Delta^{'}_{u-g}$
and  $\Delta^{'}_{g-r}$ distribution are drawn from the same parent population
are always lower that 0.1\%, but typically much lower than this. 
The lack of detection of any 
significant intrinsic {\it u-g} spread in the two remaining clusters (NGC~2419
and NGC~5466) may be due to a real lack of chemical spread but it may also
be associated with insufficient photometric accuracy and/or radial sampling. 
NGC~2419 is far more distant than any other cluster in the sample and its RGB
is observed at much fainter apparent magnitudes, implying much larger
photometric errors in the range of interest; for the same reason, the available photometry 
for this cluster samples mostly the portion of the RGB between the HB and the tip, where the amplitude of the 
UV-s effect is known to decrease.
The radial range explored here for NGC~5466 is the smallest of the whole sample, 
$r\le 4.1~r_h$.
There are indications that the UV-s effect may be weaker at very low metallicity and this may also hamper our 
ability to detect it in clusters with 
[Fe/H]$\la -1.8$.

\item In the case of M~5 ([Fe/H]=--1.34), we have demonstrated that the Na abundance
correlates with the {\it u-g} color along the RGB, in the same way as in other
clusters studied in the literature: Na-rich cluster stars have redder {\it u-g} colors
than Na-poor cluster stars of the same magnitude, while they are indistinguishable in
{\it g-r}. The same effect is not seen in M~15 ([Fe/H]=--2.32): we attribute
this to the weakening of the UV-s with decreasing metallicity, which is probably
associated with the extreme weakening of CN lines for [Fe/H]$\la -1.8$
\citep{smith02}, as CN and NH features in the region 
3000\AA $\leq\lambda\leq$ 4000\AA~are thought to be at the origin of the effect.
However, the case of M~5 illustrates that light-element abundance variations can
be traced with the ${\it u-g}$ color as done before with $U-B$, $U-V$, $U-I$,
and $c_y$ indices \citep[see, for
example][]{yong08,marino2008,han2009,car09b,car10}.

\item Dividing the RGB stars of each cluster into UV-blue and UV-red subsamples,
according to their $\Delta^{'}_{u-g}$, we found that UV-red stars are more
centrally concentrated than UV-blue stars in all the seven clusters in which we
detected a significant intrinsic UV-s. Kolmogorv-Smirnov tests have proven that the 
difference in the radial distributions of the two groups are highly significant
in all cases: the probability that UV-blue and UV-red stars be drawn
from the same radial distribution is always lower than 0.02\%. 

\item The radial profiles of the ratio of UV-red to UV-blue stars
typically show an approximately linear decline from the first sampled point (at
$\simeq 0.5-1~r_h$) out to $3-4~r_h$, where they flatten and remain
approximately flat out to the last sampled point (at $\simeq 4-8~r_h$). This
behavior is in qualitative agreement with the predictions of most recent models
of formation and chemical evolution of globular clusters. It is interesting
to note that in M~13 the profile is flat in the innermost
$\sim3~r_h$, then declines at larger radii.
The {\em shape} of
the observed profile provides a quantitative basis to test these theoretical 
scenarios, once specific models reproducing the present-day status of the
considered clusters become available. 

\end{enumerate}

The results presented here clearly suggest that the difference in the radial
distribution of first and second generation stars may be a general
characteristic of globular clusters. Moreover, it has been demonstrated that near
UV photometry can be a very efficient tool to trace light element spreads in
very large samples of RGB stars in clusters, and it is especially well suited to
studying the radial distribution of the various cluster populations. It must be
considered that the very encouraging results presented in this paper have been
obtained with a relatively small (2.5~m aperture) ground-based telescope, under
non-ideal seeing conditions \citep[$\le 1.6\arcsec$ FWHM,][]{an2008}. A survey
performed with larger telescopes and/or under good seeing conditions (e.g., with
service mode observations) and complemented with HST observations of the cluster
cores would provide a completely new insight into the problem, and would
surely contribute to shedding light on the mysterious earliest phases of the 
evolution of globular clusters.

\begin{acknowledgements} 
We are grateful to E. Vesperini for useful discussions and comments.
This research is partially supported by the PRIN-INAF 2009 
("Formation and Early Evolution of Massive Star Clusters", P.I. R. Gratton).
This research has made use of NASA's Astrophysics Data System Bibliographic Services.
This research has made use of the SIMBAD database and VizieR catalogue access tool,
operated at CDS, Strasbourg, France.\\
Funding for the SDSS and SDSS-II has been provided by the Alfred P. Sloan Foundation,
the Participating Institutions, the National Science Foundation, 
the U.S. Department of Energy, the National Aeronautics and Space Administration, 
the Japanese Monbukagakusho, the Max Planck Society, 
and the Higher Education Funding Council for England. 
The SDSS Web Site is http://www.sdss.org/.\\
The SDSS is managed by the Astrophysical Research Consortium for
the Participating Institutions. The Participating Institutions are
the American Museum of Natural History,
Astrophysical Institute Potsdam, University of Basel,
University of Cambridge, Case Western Reserve University,
University of Chicago, Drexel University, Fermilab, 
the Institute for Advanced Study, the Japan Participation Group,
 Johns Hopkins University, the Joint Institute for Nuclear Astrophysics,
 the Kavli Institute for Particle Astrophysics and Cosmology,
 the Korean Scientist Group, the Chinese Academy 
of Sciences (LAMOST), Los Alamos National Laboratory, 
the Max-Planck-Institute for Astronomy (MPIA), 
the Max-Planck-Institute for Astrophysics (MPA), 
New Mexico State University, Ohio State University, 
University of Pittsburgh, University of Portsmouth, 
Princeton University, the United States Naval Observatory, 
and the University of Washington.

\end{acknowledgements} 

\bibliographystyle{aa}
\bibliography{15662.bib}

\end{document}